\documentclass[journal]{IEEEtran}

\usepackage{tabularx}
\usepackage{times}
\usepackage{graphicx}
\usepackage{array}
\usepackage{amsmath}
\usepackage{amssymb}
\usepackage{algorithm}
\usepackage{algorithmic}
\usepackage{url}
\usepackage{multirow}
\usepackage{balance}

\usepackage[caption=false]{subfig}

\newcolumntype{V}{>{$\vcenter\bgroup\hbox\bgroup}c<{\egroup\egroup$}}
\def\Hline{\noalign{\hrule height 4\arrayrulewidth}}

\graphicspath{{./imgs/}}

\begin{document}
\title{Two Maximum Entropy Based Algorithms for Running Quantile Estimation in Non-Stationary Data Streams}

\author{Ognjen~Arandjelovi\'c,~\IEEEmembership{Member,~IEEE,}
        Duc-Son~Pham,~\IEEEmembership{Member,~IEEE,}
        and~Svetha~Venkatesh,~\IEEEmembership{Senior~Member,~IEEE}
        \thanks{O.\ Arandjelovi\'c and S. Venkatesh are with the School of Information Technology, Deakin University, Geelong, VIC, Australia (email: ognjen.arandjelovic@gmail.com). D.\ Pham is with the Department of Computing, Curtin University, PO Box U1987, WA 6845.}
        \vspace{-17pt}
        }% <-this % stops a space

\markboth{IEEE Transactions on Circuits and Systems}{}


\maketitle

\begin{abstract}
The need to estimate a particular quantile of a distribution is an important problem which frequently arises in many computer vision and signal processing applications. For example, our work was motivated by the requirements of many semi-automatic surveillance analytics systems which detect abnormalities in close-circuit television (CCTV) footage using statistical models of low-level motion features. In this paper we specifically address the problem of estimating the running quantile of a data stream when the memory for storing observations is limited. We make several major contributions: (i) we highlight the limitations of approaches previously described in the literature which make them unsuitable for non-stationary streams, (ii) we describe a novel principle for the utilization of the available storage space,  (iii) we introduce two novel algorithms which exploit the proposed principle in different ways, and (iv) we present a comprehensive evaluation and analysis of the proposed algorithms and the existing methods in the literature on both synthetic data sets and three large `real-world' streams acquired in the course of operation of an existing commercial surveillance system. Our findings convincingly demonstrate that both of the proposed methods are highly successful and vastly outperform the existing alternatives. We show that the better of the two algorithms (`data-aligned histogram') exhibits far superior performance in comparison with the previously described methods, achieving more than 10 times lower estimate errors on real-world data, even when its available working memory is an order of magnitude smaller.\\~\\
\end{abstract}

% \fi

\IEEEpeerreviewmaketitle

\section{Introduction}
\IEEEPARstart{T}{he} problem of quantile estimation is of pervasive importance across a variety of signal processing applications. It is used extensively in data mining, simulation modelling~\cite{JainChla1985}, database maintenance, risk management in finance~\cite{AdleFeldTaqq1998,SgouYaoYast2013}, and understanding computer network latencies~\cite{BuraSuri2009,CormJohnKornMuth+2004}, amongst others. A particularly challenging form of the quantile estimation problem arises when the desired quantile is high-valued (i.e.\ close to 1, corresponding to the tail of the underlying distribution) and when data needs to be processed as a stream, with limited memory capacity. An illustrative practical example of when this is the case is encountered in CCTV-based surveillance systems. In summary, as various types of low-level observations related to events in the scene of interest arrive in real-time, quantiles of the corresponding statistics for time windows of different durations are needed in order to distinguish `normal' (common) events from those which are in some sense unusual and thus require human attention. The amount of incoming data is extraordinarily large and the capabilities of the available hardware highly limited both in terms of storage capacity and processing power.

\subsection{Previous work}\label{ss:prev}
Unsurprisingly, the problem of estimating a quantile of a set has received considerable research attention, much of it in the realm of theoretical research. In particular, a substantial amount of work has focused on the study of asymptotic limits of computational complexity of quantile estimation algorithms~\cite{GuhaMcGr2009,MunrPate1980}. An important result emerging from this corpus of work is the proof by Munro and Paterson~\cite{MunrPate1980} which in summary states that the working memory requirement of any algorithm that determines the median of a set by making at most $p$ sequential passes through the input is $\Omega(n^{1/p})$ (i.e.\ asymptotically growing at least as fast as $n^{1/p}$). This implies that the exact computation of a quantile requires $\Omega(n)$ working memory. Therefore a single-pass algorithm, required to process streaming data, will necessarily produce an estimate and not be able to guarantee the exactness of its result.

Most of the quantile estimation algorithms developed for use in practice are not single-pass algorithms i.e.\ cannot be applied to streaming data \cite{GuraSriv1990}. On the other hand, many single-pass approaches focus on the exact computation of the quantile and thus, as explained previously, demand $O(n)$ storage space which is clearly an unfeasible proposition in the context we consider in the present paper. Amongst the few methods described in the literature and which satisfy our constraints are the histogram-based method of Schmeiser and Deutsch~\cite{SchmDeut1977} (with a similar approach described by McDermott \textit{et al.} \cite{McDeBabuLiecLin2007}), and the $P^2$ algorithm of Jain and Chlamtac~\cite{JainChla1985}. Schmeiser and Deutsch maintain a preset number of bins, scaling their boundaries to cover the entire data range as needed and keeping them equidistant. Jain and Chlamtac attempt to maintain a small set of \textit{ad hoc} selected key points of the data distribution, updating their values using quadratic interpolation as new data arrives. Lastly, random sample methods, such as that described by Vitter~\cite{Vitt1985}, and Cormode and Muthukrishnan~\cite{CormMuth2005}, use different sampling strategies to fill the available buffer with random data points from the stream, and estimate the quantile using the distribution of values in the buffer.

In addition to the \textit{ad hoc} elements of the previous algorithms for quantile estimation on streaming data, which itself is a sufficient cause for concern when the algorithms need to be deployed in applications which demand high robustness and well understood failure modes, it is also important to recognize that an implicit assumption underlying these approaches is that the data is governed by a stationary stochastic process. The assumption is often invalidated in real-world applications. As we will demonstrate in Section~\ref{s:eval}, a consequence of this discrepancy between the model underlying existing algorithms and the nature of data in practice is a major deterioration in the quality of quantile estimates. Our principal aim is thus to formulate a method which can cope with non-stationary streaming data in a more robust manner.

\section{Proposed algorithms}
We start this section by formalizing the notion of a quantile. This is then followed by the introduction of the key premise of our contribution and finally a description of two algorithms which exploit the underlying idea in different ways~\cite{AranPhamVenk2014}. The algorithms are evaluated on real-world data in the next section.

\subsection{Quantiles}
Let $p$ be the probability density function of a real-valued random variable $X$. Then the $q$-quantile $v_q$ of $p$ is defined as:
\begin{align}
  \int_{-\infty}^{v_q} p(x)~dx = q.
\end{align}
Similarly, the $q$-quantile of a finite set $D$ can be defined as:
\begin{align}
  \left|\{ x~:~ x \in D \text{ and  } x \leq v_q\}\right| \leq q \times |D|.
\end{align}
In other words, the $q$-quantile is the smallest value below which $q$ fraction of the total values in a set lie~\cite{Wilc2012}. The concept of a quantile is thus intimately related to the tail behaviour of a distribution.

\subsection{Methodology: maximum entropy histograms}\label{ss:maxEnt}
\begin{figure}[htb]
  \centering
  \includegraphics[width=0.45\textwidth]{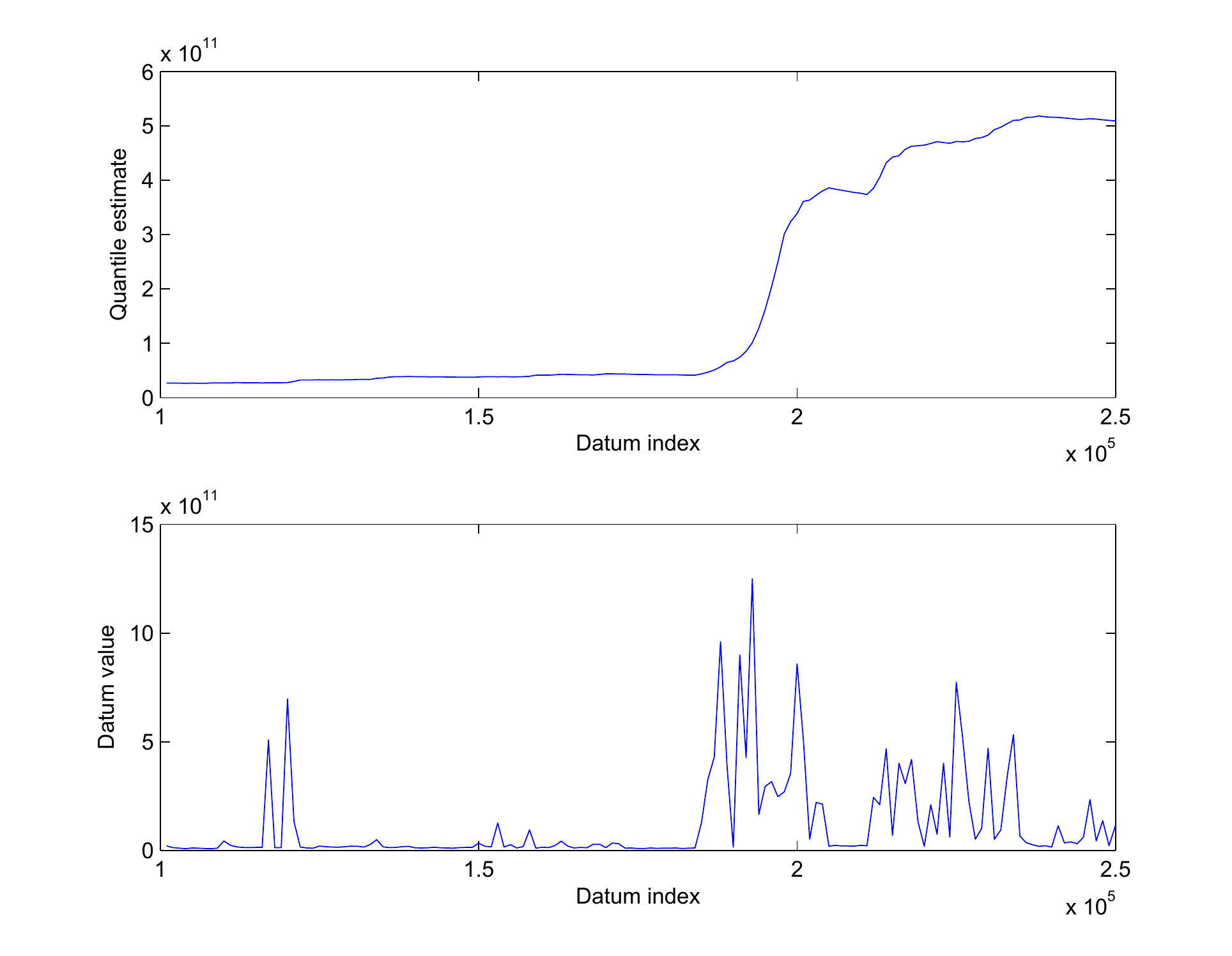}
  \caption{An example of a rapid change in the value of a quantile (specifically the 0.95-quantile in this case) on a real-world data stream. }
  \label{f:change}
\end{figure}

A consequence of the non-stationarity of data streams that we are dealing with is that at no point in time can it be assumed that the historical distribution of data values is representative of its future distribution. This is true regardless of how much historical data has been seen. Thus, the value of a particular quantile can change greatly and rapidly, in either direction (i.e.\ increase or decrease). This is illustrated on an example, extracted from a real-world data set used for surveillance video analysis (the full data corpus is used for comprehensive evaluation of different methods in Section~\ref{s:eval}), in Figure~\ref{f:change}. In particular, the top plot in this figure shows the variation of the ground-truth 0.95-quantile which corresponds to the data stream shown in the bottom plot. Notice that the quantile exhibits little variation over the course of approximately the first 75\% of the duration of the time window (the first 190,000 data points). This corresponds to a period of little activity in the video from which the data is extracted (see Section~\ref{s:eval} for a detailed explanation). Then, the value of the quantile increases rapidly for over an order of magnitude -- this is caused by a sudden burst of activity in the surveillance video and the corresponding change in the statistical behaviour of the data.

To be able to adapt to such unpredictable variability in input it is therefore not possible to focus on only a part of the historical data distribution but rather it is necessary to store a `snapshot' of the entire distribution. We achieve this using a histogram of a fixed length, determined by the available working memory. In contrast to the previous work which either distributes the bin boundaries equidistantly or uses \textit{ad hoc} adjustments, our idea is to maintain bins in a manner which maximizes the entropy of the corresponding estimate of the historical data distribution.

Recall that the entropy of a discrete random variable is:
\begin{align}
  H_r=-\sum_{x \in X} p(x) \log_2 p(x)
\end{align}
where $p(x)$ is the probability of the random variable taking on the value $x$, and $X$ the set of all possible values of $x$~\cite{Shan1948} (of course $\sum_{x \in X} p(x)=1$). Therefore, the entropy of a histogram with $n$ bins and the corresponding bin counts $c_1, c_2,\ldots, c_n$ is:
\begin{align}
  H_h=-\sum_{i=1}^n \frac{c_i}{C} \log_2 \left( \frac{c_i}{C} \right)
\end{align}
where $C=\sum_{i=1}^n c_i$ is the normalization factor which makes $c_i/C$ the probability of a randomly selected datum belonging to the $i$-th bin~\cite{WangYe2005}. Our goal is to dynamically allocate and adjust bin boundaries in a manner which maximizes the associated entropy for a fixed number of bins $n$ (which is an input parameter whose value is determined by practical constraints). Much like the problem of computing a specific quantile which, as we discussed with reference to Munro and Paterson's work~\cite{MunrPate1980} in Section~\ref{ss:prev}, is not solvable exactly using a single-pass algorithm, the construction of the maximal entropy histogram as defined above is not possible to guarantee in the setting adopted in this paper whereby only a limited and fixed amount of storage is available, and historical data is inaccessible.

\subsection{Method 1: interpolated bins}\label{ss:method1}
The first method based around the idea of maximum entropy bins we introduce in this paper
readjusts the boundaries of a fixed number of bins after the arrival of each new data point $d_{i+1}$. Without loss of generality let us assume that each datum is positive i.e.\ that $d_i > 0$. Furthermore, let the upper bin boundaries before the arrival of $d_i$ be $b^i_1, b^i_2,\ldots, b^i_n$, where $n$ is the number of available bins. Thus, the $j$-th bin's catchment range is $(b^i_{j-1}, b^i_j]$ where we will take that $b^i_0=0$ for all $i$. We wish to maintain the condition that the piece-wise uniform probability density function approximation of the historical data distribution described by this histogram has the maximum entropy of all those possible with the histogram of the same length. This is achieved by having equiprobable bins. Thus, before the arrival of $d_{i+1}$, the number of historical data points in each bin is the same and equal to $i/n$. The corresponding cumulative density is given by:
\begin{align}
  f^i(d)=\frac{1}{n} \times \left[j + \frac{d-b^i_{j-1}}{b^i_j-b^i_{j-1}} \right]
\end{align}
and
\begin{align}
  b^i_{j-1} < d \leq b^i_j.
\end{align}

After the arrival of $d_i$, but before the readjustment of bin boundaries, the cumulative density becomes:
\begin{align}
  \widetilde{f}^i(d)= \begin{cases}
            \frac{i}{i+1} \times \frac{1}{n} \times \left[j + \frac{d-b^i_{j-1}}{b^i_j-b^i_{j-1}} \right] ~~~~~~~~~~~~~~\text{ for } d < d_i\\
            \frac{i}{i+1} \times \frac{1}{n} \times \left[j + \frac{d-b^i_{j-1}}{b^i_j-b^i_{j-1}} \right] +\frac{1}{i+1} ~~~~\text{ for } d \geq d_i\\
          \end{cases}
\end{align}
Lastly, to maintain the invariant of equiprobable bins, the bin boundaries are readjusted by linear interpolation of the corresponding inverse distribution function.

\subsubsection{Initialization} The initialization of the proposed algorithm is simple. Specifically, until the buffer is filled, i.e.\ until the number of unique stream data points processed exceeds $n$, the maximal entropy histogram is constructed by allocating each unique data value its own bin. This is readily achieved by making each unique data value the top boundary of a bin.

\subsection{Method 2: data-aligned bins}\label{ss:method2}
The algorithm described in the preceding section appears optimal in that it always attempts to realign bins so as to maintain maximum entropy of the corresponding approximation for the given size of the histogram. However, a potential source of errors can emerge cumulatively as a consequence of repeated interpolation, done after every new datum. Indeed, we will show this to be the case empirically in Section~\ref{s:eval}. We now introduce an alternative approach which aims to strike a balance between some unavoidable loss of information, inherently a consequence of the need to readjust an approximation of the distribution of a continually growing data set, and the desire to maximize the entropy of this approximation.

Much like in the previous section, bin boundaries are potentially altered each time a new datum arrives. There are two main differences in how this is performed. Firstly, unlike in the previous case, bin boundaries are not allowed to assume arbitrary values; rather, their values are constrained to the values of the seen data points. Secondly, only at most a single boundary is adjusted for each new datum. We now explain this process in detail.

\begin{figure}[htb]
  \centering
  \includegraphics[width=0.4\textwidth]{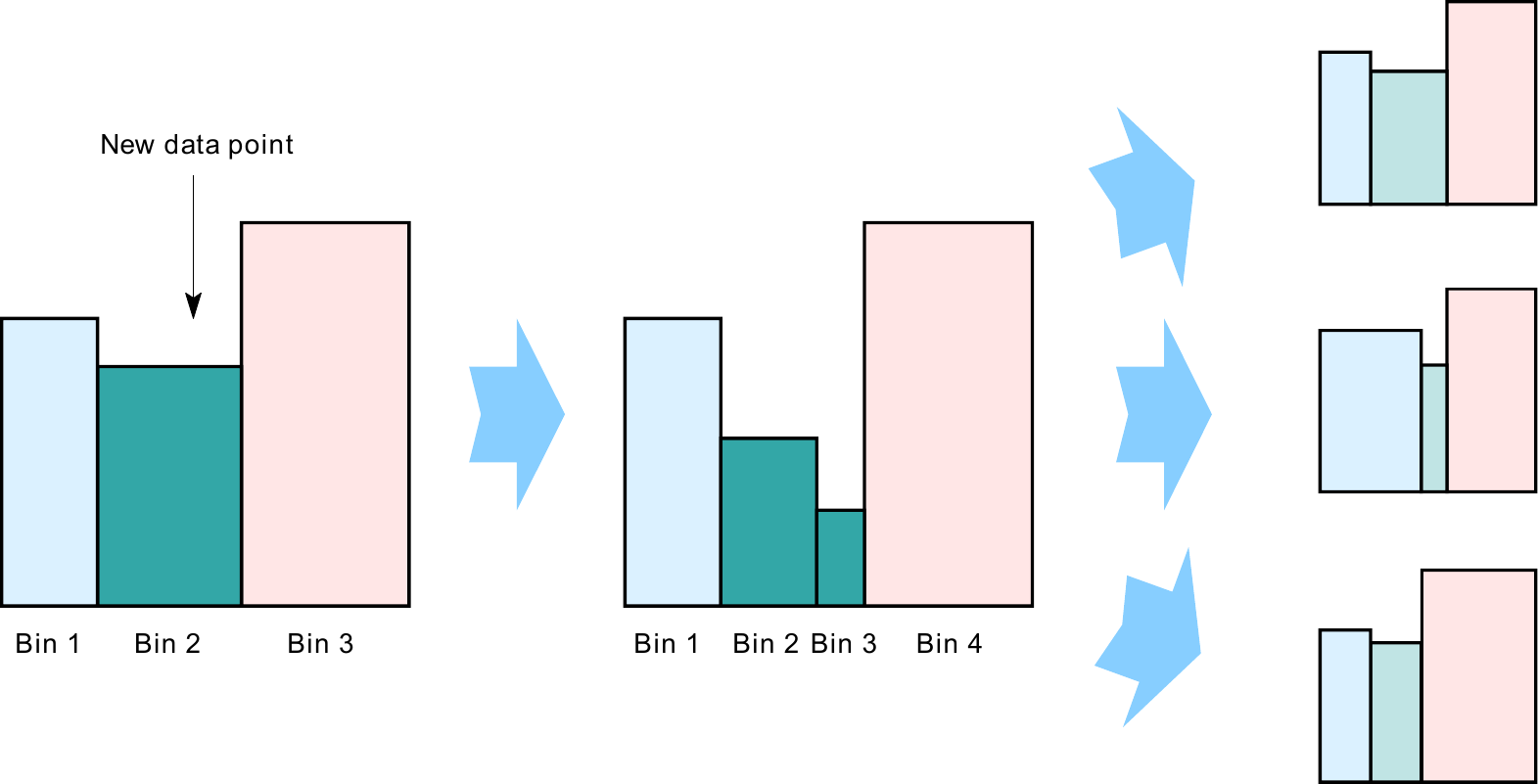}
  \caption{ Conceptual illustration of the key update step of the second algorithm first described in the present paper. The algorithm attempts to maximize the entropy of the histogram approximating the distribution of historical data while using bins with data-aligned boundaries. The figure shows the initial histogram before the arrival of the new datum whose value is indicated by the arrow (left-most illustration). A temporary new bin is created with the boundary coinciding with the value of the new datum (shown in the centre of the diagram). Lastly, to maintain a fixed number of bins, a single merging of two neighbouring bins is performed; the three small illustrations on the right-hand side show the possible options. The merge which results in highest entropy is chosen and actually performed. }
  \label{f:merge}
\end{figure}

As before, let the upper bin boundaries before the arrival of a new data point be $b^i_1, b^i_2,\ldots, b^i_n$. Since unlike in the case of the previous algorithm in general the bins will not be equiprobable we also have to maintain a corresponding list $c^i_1, c^i_2,\ldots, c^i_n$ which specifies the corresponding data counts. Each time a new data point arrives, an $(n+1)$-st bin is created temporarily. If the value of the new datum is greater than $b^i_n$ (and thus greater than any of the historical data), a new bin is created after the current $n$-th bin, with the upper boundary set at $d(i)$. The corresponding datum count $c$ of the bin is set to 1. Alternatively, if the value of the new data point is lower than $b^i_n$ then there exists $j$ such that:
\begin{align}
  b^i_{j-1} < d \leq b^i_j,
\end{align}
and the new bin is inserted between the $(j-1)$-st and $j$-th bin. Its datum count is estimated using linear interpolation in the following manner:
\begin{align}
  c = c^i_j \times \frac{d-b^i_{j-1}}{b^i_j-b^i_{j-1}} + 1.
\end{align}
Thus, regardless of the value of the new data point, temporarily the number of bins is increased by 1. The original number of bins is then restored by merging exactly a single pair of neighbouring bins. For example, if the $k$-th and $(k+1)$-st bin are merged, the new bin has the upper boundary value set to the upper boundary value of the former $(k+1)$-st bin, i.e.\ $b^i_{k+1}$, and its datum count becomes the sum of counts for the $k$-th and $(k+1)$-st bins, i.e.\ $c^i_k + c^i_{k+1}$.  The choice of which neighbouring pair to merge, out of $n$ possible options, is made according to the principle stated in Section~\ref{ss:maxEnt}, i.e.\ the merge actually performed should maximize the entropy of the new $n$-bin histogram. This is illustrated conceptually in Figure~\ref{f:merge}.

\subsubsection{Initialization} The initialization of the histogram in the proposed method can be achieved in the same manner as in the interpolated bins algorithm introduced previously. To repeat, until the buffer is filled, i.e.\ until the number of unique stream data points processed exceeds $n$, the maximal entropy histogram is constructed by making each unique data value the top boundary of a bin, thereby allocating each unique data value its own bin.

\section{Evaluation and results}\label{s:eval}
\begin{figure*}[htb]
  \centering
  \subfloat[]{\includegraphics[width=0.3\textwidth]{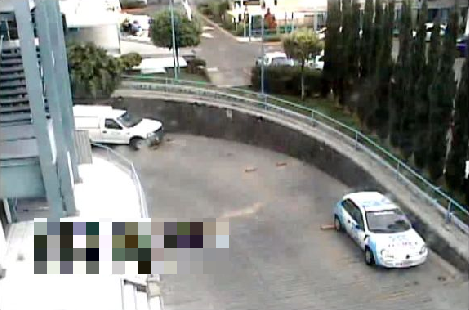}}~~~~~
  \subfloat[]{\includegraphics[width=0.3\textwidth]{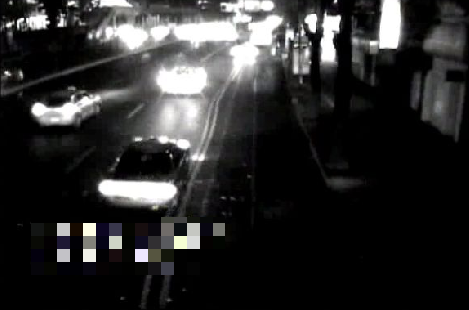}}~~~~~
  \subfloat[]{\includegraphics[width=0.3\textwidth]{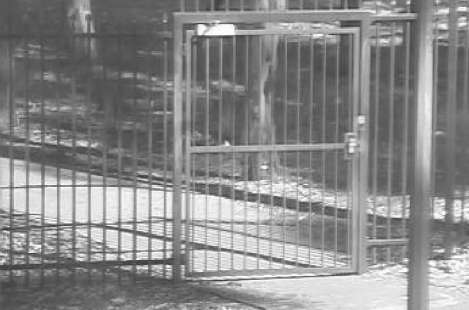}}
  \caption{ Screenshots of the three scenes used to acquire the data used in our experiments. Note that these are real, operational CCTV cameras, which were not specifically installed for the purpose of data acquisition for the present work. Also see Figure~\ref{f:streams}. }
  \label{f:scenes}
\end{figure*}

We now turn our attention to the evaluation of the proposed algorithms. In particular, to assess their effectiveness and compare them with the algorithms described in the literature (see Section~\ref{ss:prev}), in this section we report their performance on two synthetic data sets and three large `real-world' data streams. Our aim is first to use simple synthetic data to study the algorithms in a well understood and controlled setting, before applying them on corpora collected by systems deployed in practice. Specifically, the `real-world' streams correspond to motion statistics used by an existing CCTV surveillance system for the detection of abnormalities in video footage. It is important to emphasize that the data we used was not acquired for the purpose of the present work nor were the cameras installed with the same intention. Rather, we used data which was acquired using existing, operational surveillance systems. In particular, our data comes from three CCTV cameras, two of which are located in Mexico and one in Australia. The scenes they overlook are illustrated using a single representative frame per camera in Figure~\ref{f:scenes}. Table~\ref{t:data} provides a summary of some of the key statistics of the three data sets. We explain the source of these streams and the nature of the phenomena they represent in further detail in Section~\ref{sss:dataReal}.

\begin{table}[htb]
  \caption{Key statistics of the three real-world data sets used in our evaluation. These were acquired using three existing CCTV cameras in operation in Australia and Mexico.
  }
  \renewcommand{\arraystretch}{1.4}
  \vspace{0pt}
  \centering
  \small
  \begin{tabular}{l|ccc}
  \Hline
  Data set  & Data points & Mean value & Standard deviation\\
  \hline
  Stream~1  &    $555,022$ & $7.81\times 10^{10}$ & $1.65\times 10^{11}$\\
  Stream~2  & $10,424,756$ & $2.25$               & $15.92$ \\
  Stream~3  &  $1,489,618$ & $1.51\times 10^{5}$  & $2.66\times 10^{6}$\\
  \Hline
  \end{tabular}
  \label{t:data}
\end{table}

\subsection{Evaluation data}

\subsubsection{Synthetic data}
The first synthetic data set that we used for the evaluation in this paper is a simple stream $x_1,x_2,\ldots,x_{n_1}$ generated by drawing each datum $x_i$ independently from a normal distribution represented by the random variable $X$:
\begin{align}
  X \sim \mathcal{N}(5,1)
\end{align}
Therefore, this sequence has a stationary distribution. We used $n_1=1,000,000$ data points.

The second synthetic data set is somewhat more complex. Specifically, each datum $y_i$ in the stream $y_1,y_2,\ldots,y_{n_1}$ is generated as follows:
\begin{align}
  y_i=c_i \times y^{(1)}_i + (1-c_i) \times y^{(2)}_i
\end{align}
where $c_i$ is drawn from a discrete uniform distribution over the set $\left\{ 0,1 \right\}$, and $y^{(1)}_i$ and $y^{(2)}_i$ from normal distributions represented by the random variables $Y_1$ and $Y_2$ respectively:
\begin{align}
  Y_1& \sim \mathcal{N}(5,1)\\
  Y_2& \sim \mathcal{N}(10,4).
\end{align}
In intuitive terms, a datum is generated by flipping a fair coin and then depending on the outcome drawing the value either from $Y_1$ or $Y_2$. Notice that this data set therefore does not have the property of stationarity. As in the first experiment we used $n_2=1,000,000$ data points.

\subsubsection{Real-world surveillance data}\label{sss:dataReal}
\begin{figure}[htb]
  \centering
  \subfloat[Data stream~1]{\includegraphics[width=0.407\textwidth]{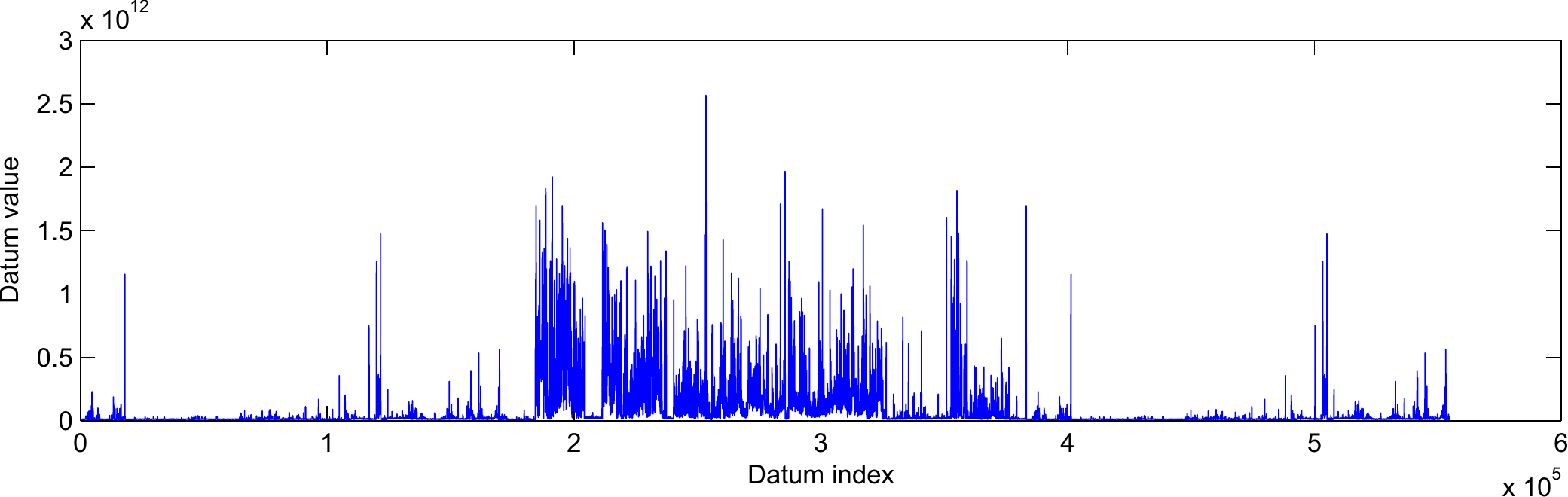}\label{f:stream1}}\\
  \subfloat[Data stream~2]{\includegraphics[width=0.495\textwidth]{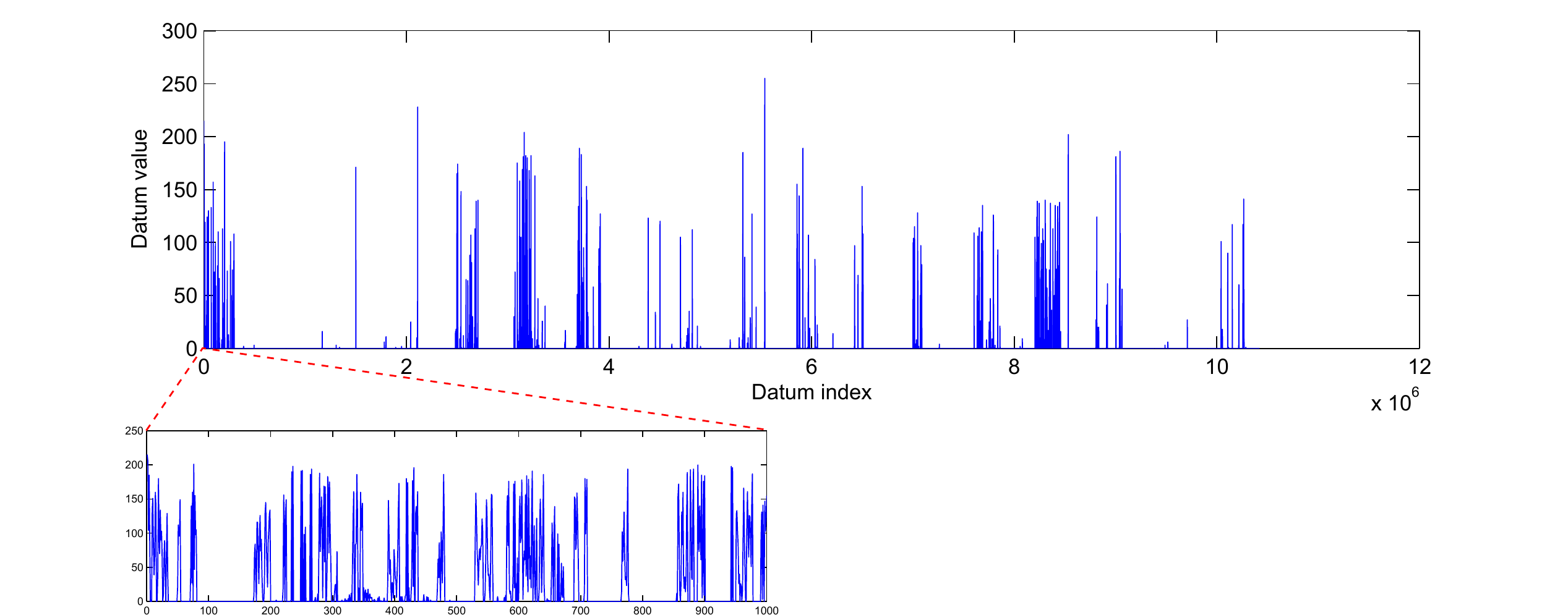}}\\
  \subfloat[Data stream~3]{\includegraphics[width=0.495\textwidth]{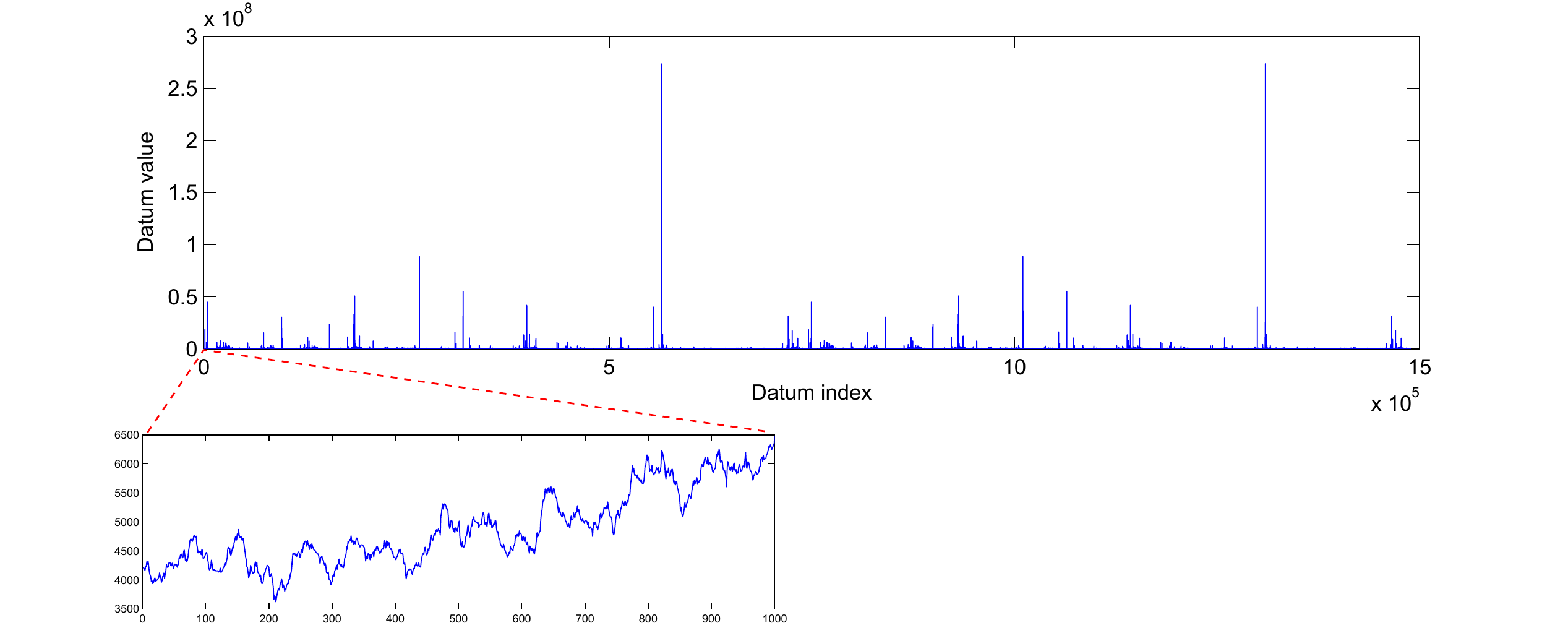}}
  \caption{ The three large data streams used to evaluate the performance of the proposed algorithms and compare them with the approaches previously described in the literature. Also see Figure~\ref{f:scenes}. }
  \label{f:streams}
\end{figure}

Computer-assisted video surveillance data analysis is of major commercial and law enforcement interest. On a broad scale, systems currently available on the market can be grouped into two categories in terms of their approach. The first group focuses on a relatively small, predefined and well understood subset of events or behaviours of interest such as the detection of unattended baggage, violent behaviour, etc \cite{Phil,LaveKhanThur2007}. The narrow focus of these systems prohibits their applicability in less constrained environments in which a more general capability is required. In addition, these approaches tend to be computationally expensive and error prone, often requiring fine tuning by skilled technicians. This is not practical in many circumstances, for example when hundreds of cameras need to be deployed as often the case with CCTV systems operated by municipal authorities. The second group of systems approaches the problem of detecting suspicious events at a semantically lower level~\cite{iCet,Aran2011a,Inte,MartAran2010,PhamAranVenk2014}. Their central paradigm is that an unusual behaviour at a high semantic level will be associated with statistically unusual patterns (also `behaviour' in a sense) at a low semantic level -- the level of elementary image/video features. Thus methods of this group detect events of interest by learning the scope of normal variability of low-level patterns and alerting to anything that does not conform to this model of what is expected in a scene, without `understanding' or interpreting the nature of the event itself. These methods uniformly start with the same procedure for feature extraction. As video data is acquired, firstly a dense optical flow field is computed. Then, to reduce the amount of data that needs to be processed, stored, or transmitted, a thresholding operation is performed. This results in a sparse optical flow field whereby only those flow vectors whose magnitude exceeds a certain value are retained; non-maximum suppression is applied here as well. Normal variability within a scene and subsequent novelty detection are achieved using various statistics computed over this data. The three data streams, shown partially in Figure~\ref{f:streams}, correspond to the values of these statistics (their exact meaning is proprietary and has not been made known fully to the authors of the present paper either; nonetheless we have obtained permission to make the data public as we shall do following the acceptance of the paper). Observe the non-stationary nature of the data streams which is evident both on the long and short time scales (magnifications are shown for additional clarity and insight).

\subsection{Results}
We now compare the performance of our algorithms with the three alternatives from the literature described in Section~\ref{ss:prev}: (i) the $P^2$ algorithm of Jain and Chlamtac~\cite{JainChla1985}, (ii) the random sample based algorithm of Vitter~\cite{Vitt1985}, and (iii) the uniform adjustable histogram of Schmeiser and Deutsch~\cite{SchmDeut1977}.

\subsubsection{Synthetic data}\label{sss:resSynth}
\begin{figure*}[htb]
  \centering
  \subfloat[0.95 quantile]{\includegraphics[width=0.33\textwidth]{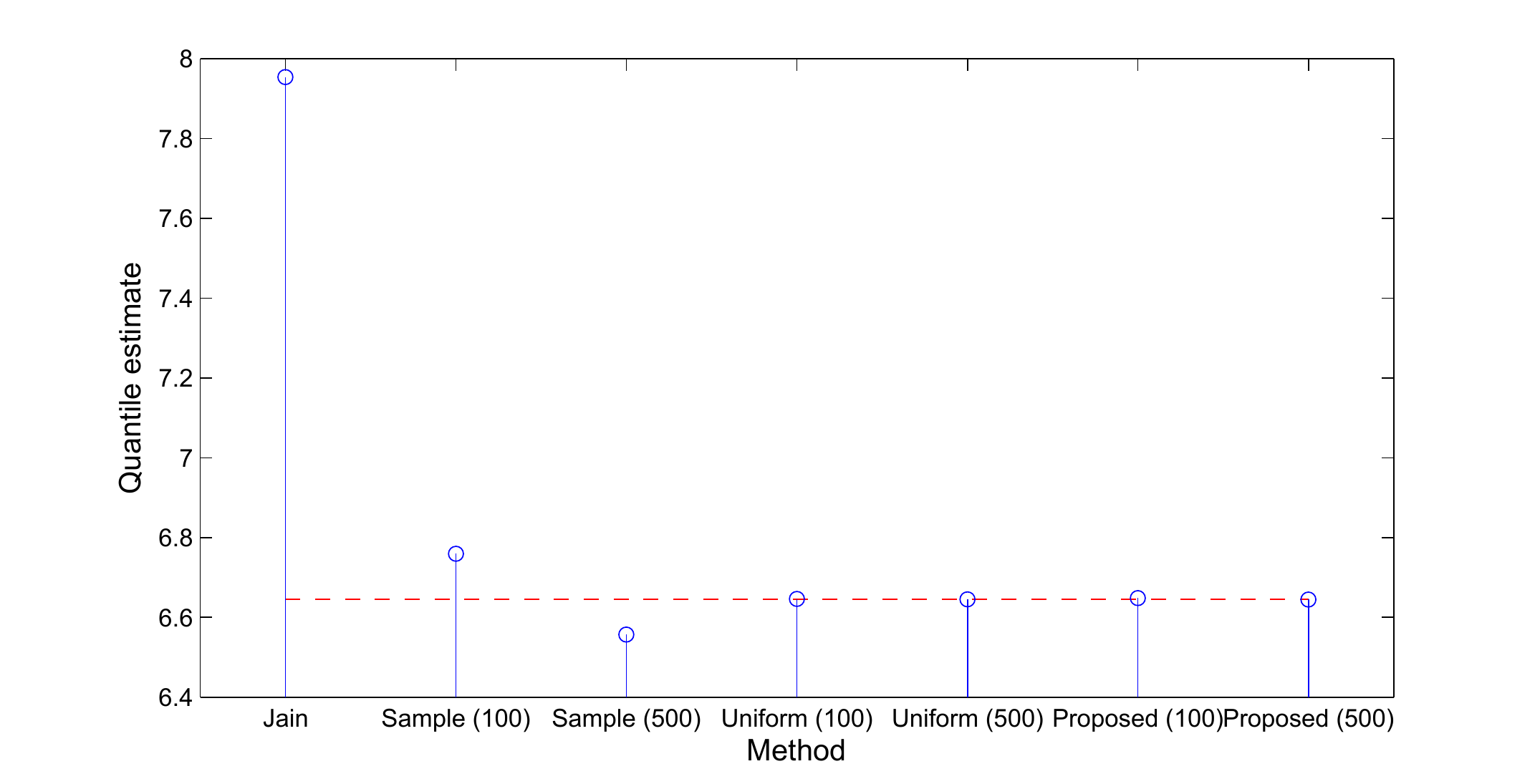}}
  \subfloat[0.99 quantile]{\includegraphics[width=0.33\textwidth]{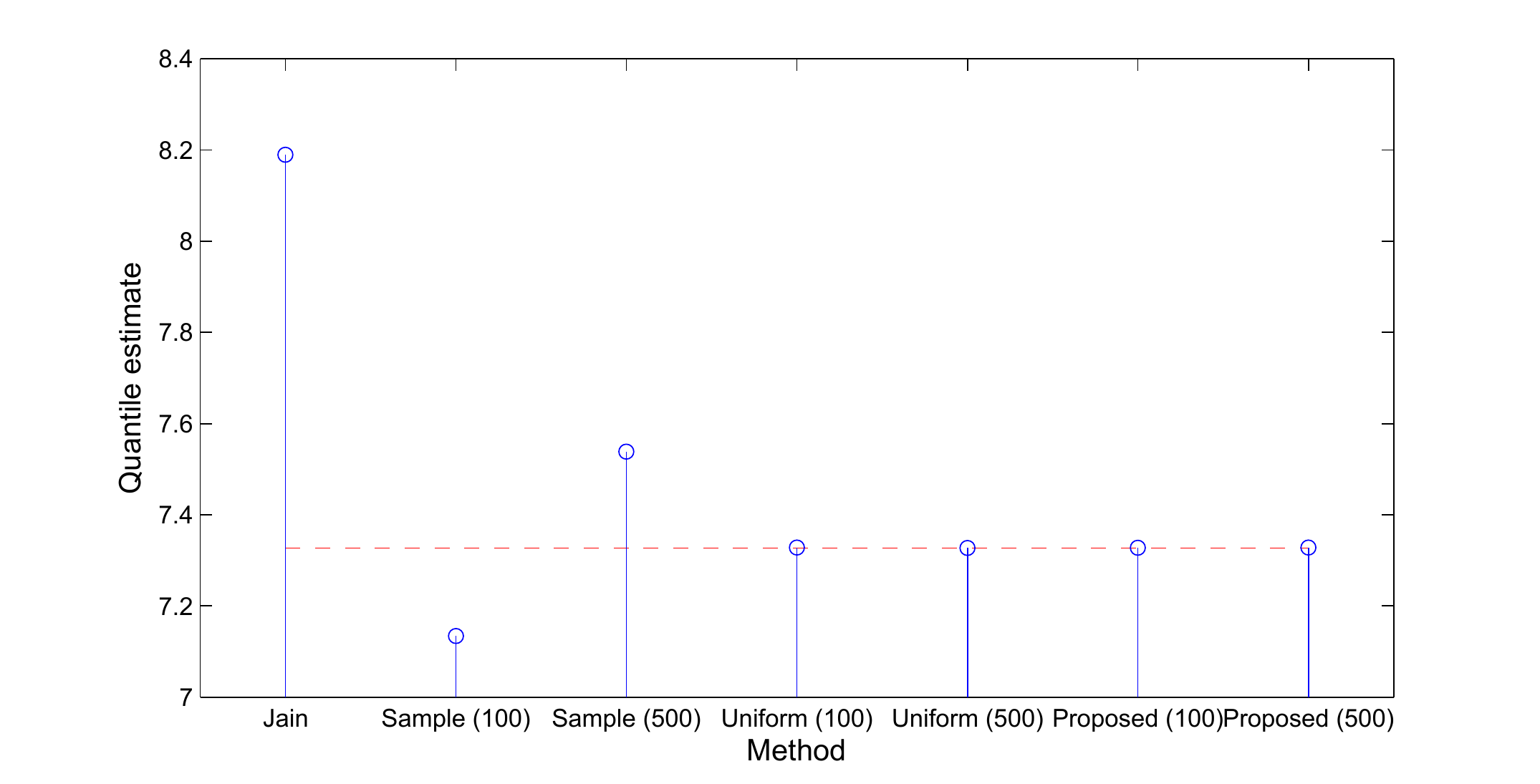}}
  \subfloat[0.995 quantile]{\includegraphics[width=0.33\textwidth]{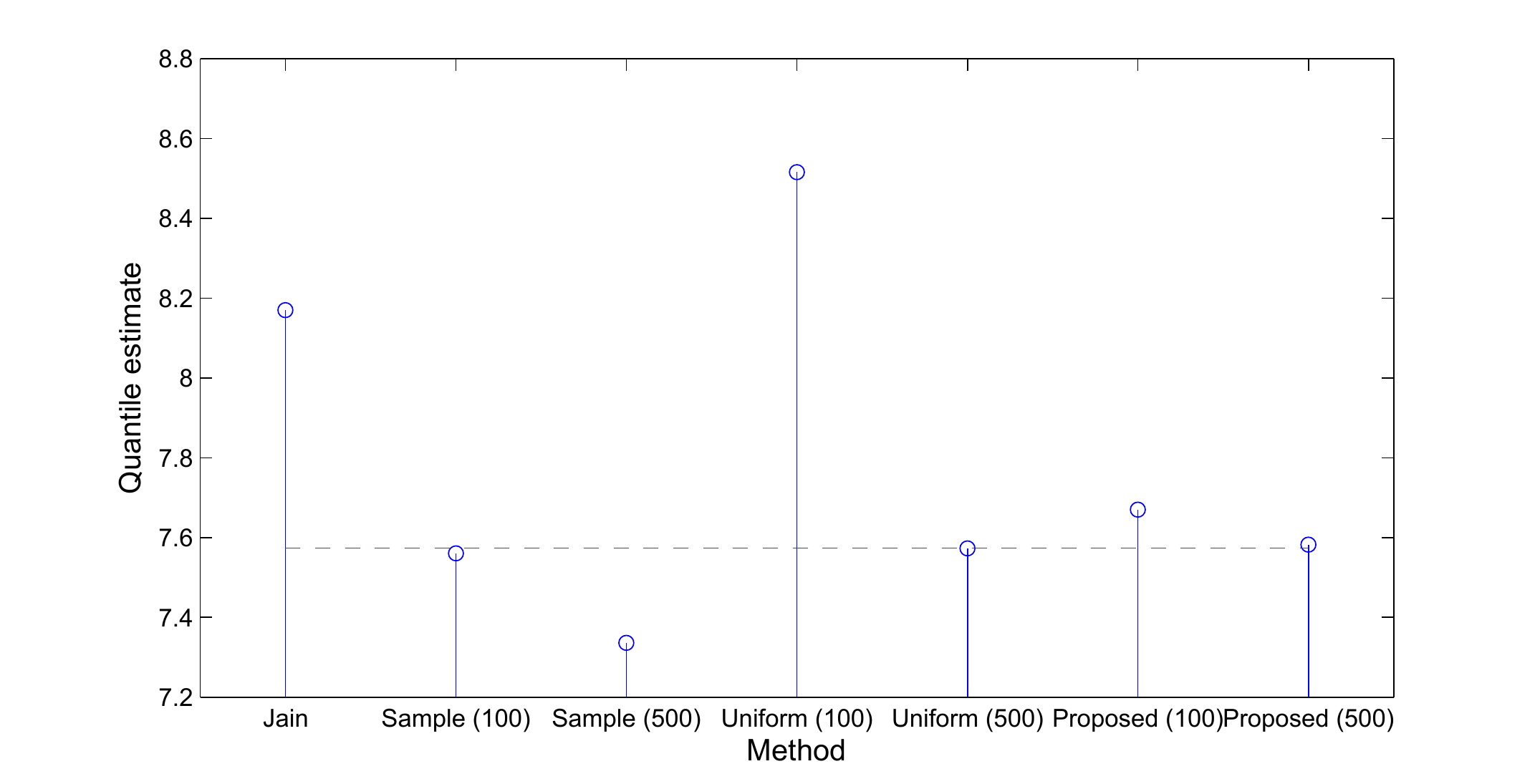}}
  \caption{ A comparison of different methods on the first synthetic data set used in this paper. This stream has stationary statistical characteristics and was generated by drawing each datum (of 1,000,000 in total) independently from the normal distribution $\mathcal{N}(5,1)$. The label `Jain' refers to the $P^2$ algorithm of Jain and Chlamtac~\cite{JainChla1985}, `Sample' to the random sample based algorithm of Vitter~\cite{Vitt1985}, `Uniform' to the uniform adjustable histogram of Schmeiser and Deutsch~\cite{SchmDeut1977}, and `Proposed' to the data-aligned bins described in Section~\ref{ss:method2}. The number in brackets after a method name signifies the size of its buffer i.e.\ available working memory. The dotted red line shows the true quantile values. }
  \label{f:res_synth1}
\end{figure*}

We start by examining the results of different algorithms on the first and simplest synthetic stream, with stationary characteristics and data drawn from a normal distribution. Different estimates for the quantile values of 0.95, 0.99, and 0.995 are shown in the stem plots of Figure~\ref{f:res_synth1}. Several trends are immediately obvious. Firstly, Jain and Chlamtac's algorithm consistently performed worse, significantly so, than all other methods in all three experiments.  This is unsurprising, given that the algorithm uses the least amount of memory. The best performance across all experiments was exhibited by the data-aligned algorithm introduced in this paper, while the relative performances of the sample-based algorithm of Vitter and the uniform histogram method are not immediately clear, one performing better than the other in some cases and \emph{vice versa} in others.

\begin{figure}[htb]
  \centering
  \includegraphics[width=0.48\textwidth]{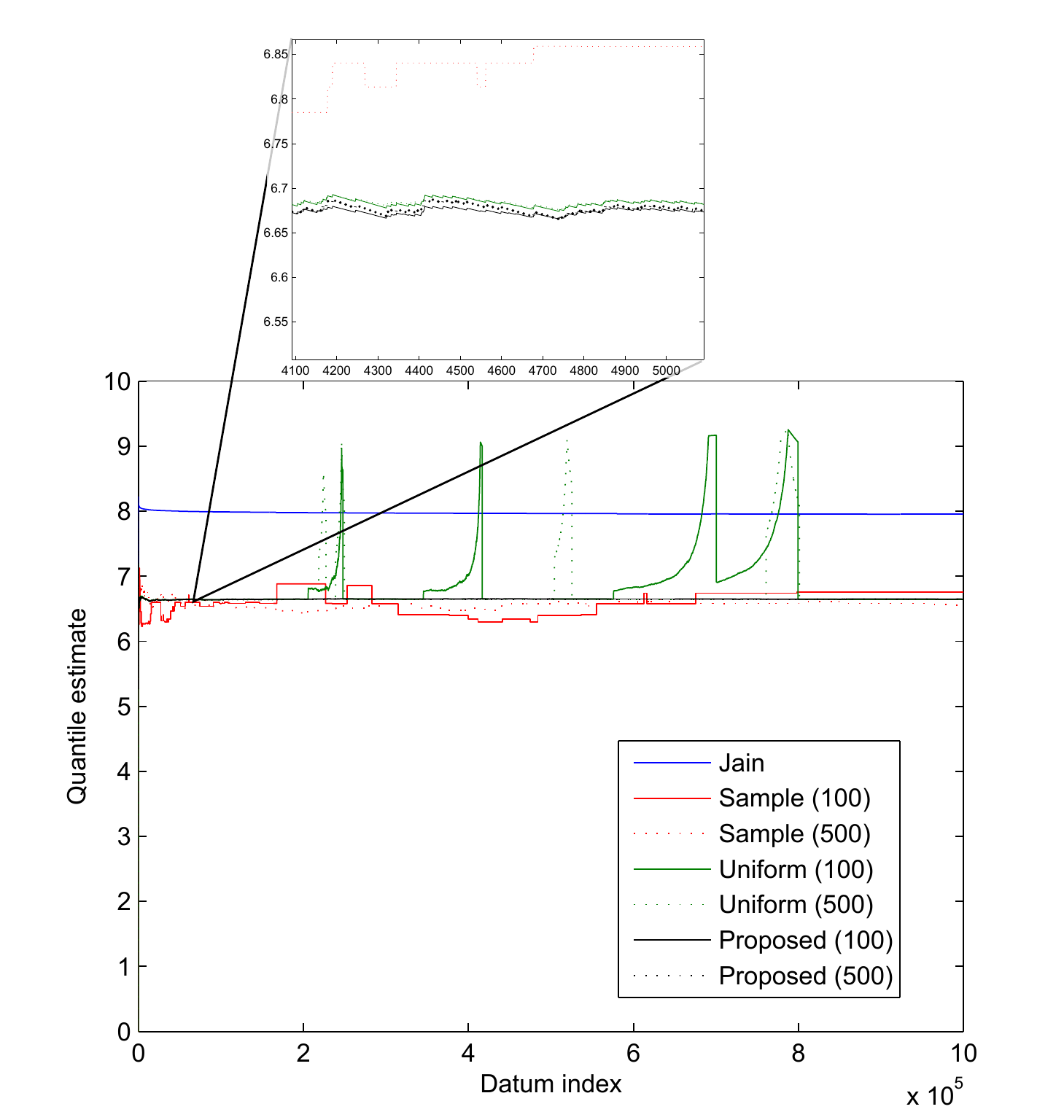}
  \caption{ Running estimate of the 0.95-quantile produced by different methods on our first synthetic data set. The label `Jain' refers to the $P^2$ algorithm of Jain and Chlamtac~\cite{JainChla1985}, `Sample' to the random sample based algorithm of Vitter~\cite{Vitt1985}, `Uniform' to the uniform adjustable histogram of Schmeiser and Deutsch~\cite{SchmDeut1977}, and `Proposed' to the data-aligned bins described in Section~\ref{ss:method2}. The number in brackets after a method name signifies the size of its buffer i.e.\ available working memory. Both of the proposed algorithms rapidly achieve high accuracy which is maintained throughout. Since the two running estimates are indistinguishable by the naked eye at the scale of the original plot, for the benefit of the reader a small portion of the plot under the magnification of approximately 200 times is shown too; at this scale a small difference between the proposed methods can be observed.  }
  \label{f:res_synth1_plot}
\end{figure}

Figure~\ref{f:res_synth1} also shows that in all cases except that of the sample-based algorithm of Vitter in the estimation of 0.995-quantile, a particular method performed better when its available storage space was increased. This observation too is in line with theoretical expectations. However, this is only a partial picture because it offers only a snapshot of the estimates after all data has been processed. The plot in Figure~\ref{f:res_synth1_plot} plots the running estimates of all algorithms as more and more data is seen, and reveals further insight. The data-aligned bins algorithm proposed herein (black lines) can again be seen to perform the best, showing little fluctuation during the processing of the stream, its performance with 100 being no worse than with 500 bins. The plot also confirms the inferiority of Jain and Chlamtac's method. A more interesting result pertains to the comparison of the sample-based algorithm of Vitter and the uniform adjustable histogram of Schmeiser and Deutsch. Specifically, despite its good accuracy at most times, the latter can be seen to suffer intermittently from large errors, as witnessed by the pronounced high-frequency strays in the plot (green lines). These can be readily explained by considering the operation of the algorithm and in particular its behaviour when a new extreme datum, outside of the range of the current uniform histogram, arrives. In such instances, the bin boundaries need to be readjusted and are consequently greatly altered, producing a number of poorly-sampled bins. This results in inaccurate estimates (most markedly of high quantiles), which are transient in nature as the sampling becomes more accurate with the arrival of further data.

\begin{figure*}[htb]
  \centering
  \subfloat[Accuracy: 0.01 (1\%)]{\includegraphics[width=0.33\textwidth]{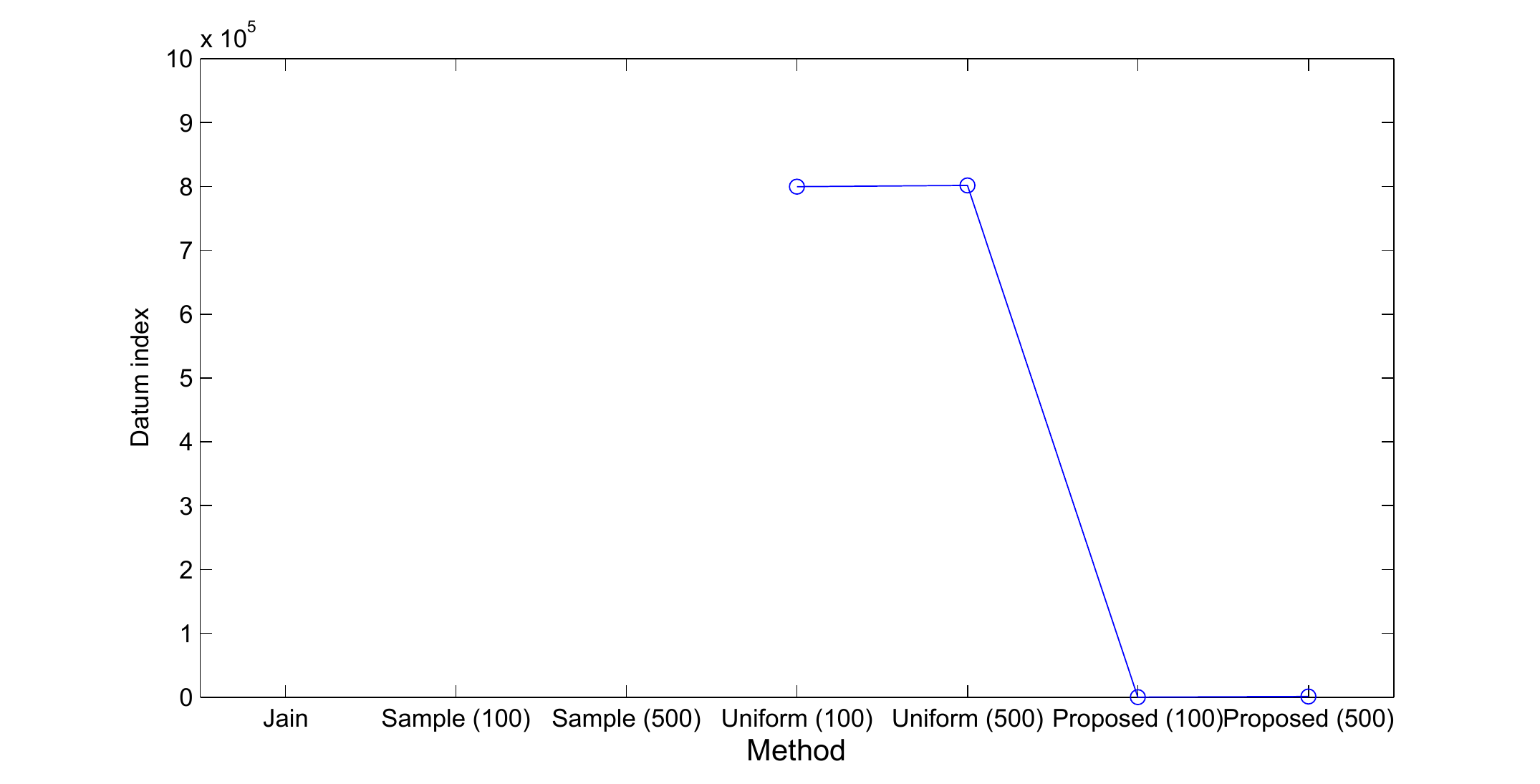}}
  \subfloat[Accuracy: 0.05 (5\%)]{\includegraphics[width=0.33\textwidth]{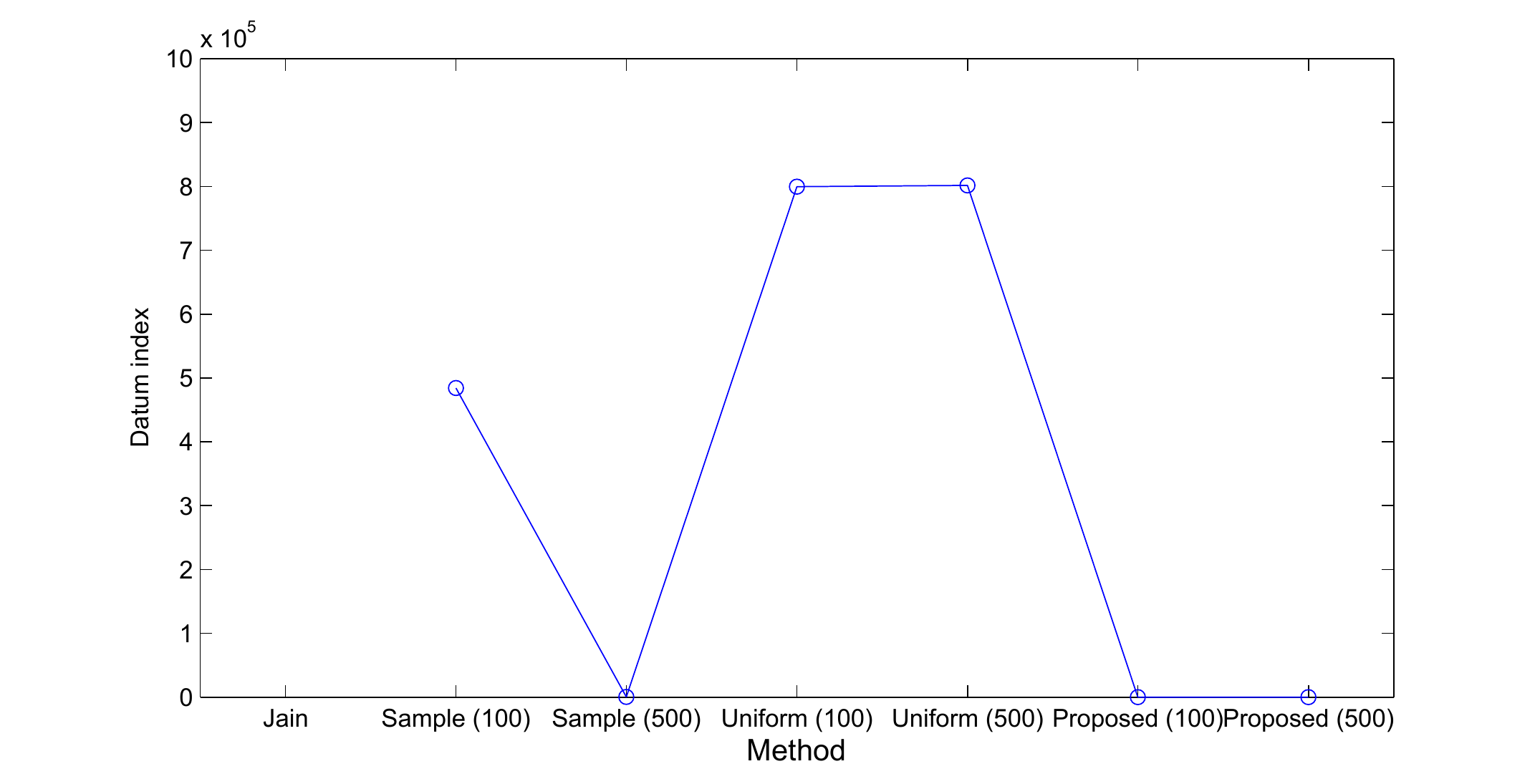}}
  \subfloat[Accuracy: 0.1 (10\%)]{\includegraphics[width=0.33\textwidth]{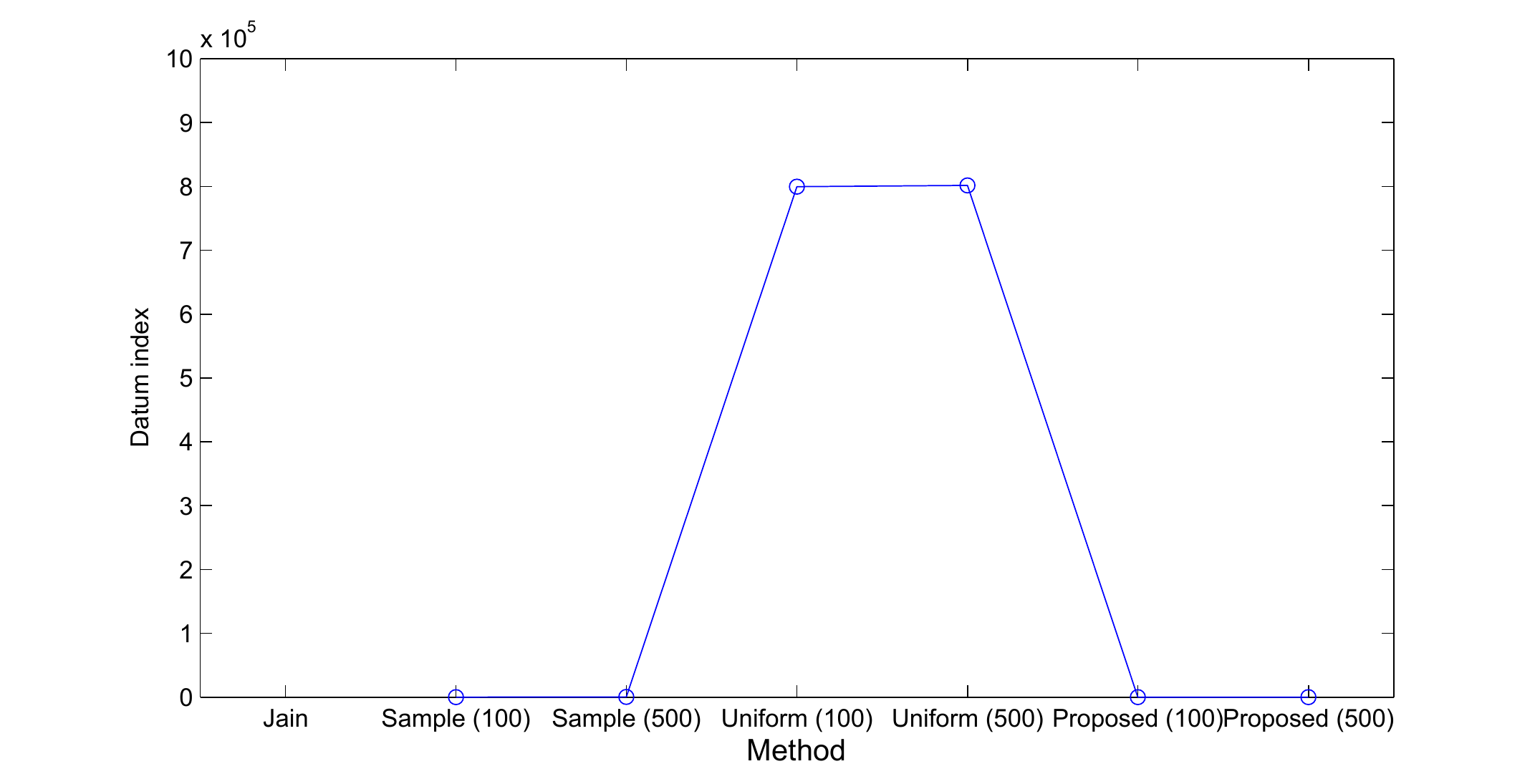}}
  \caption{ Running time (as number of historical data points) until the relative error of the estimate of an algorithm on a stream permanently drops to at most the particular accuracy (see main text for formal definition). The label `Jain' refers to the $P^2$ algorithm of Jain and Chlamtac~\cite{JainChla1985}, `Sample' to the random sample based algorithm of Vitter~\cite{Vitt1985}, `Uniform' to the uniform adjustable histogram of Schmeiser and Deutsch~\cite{SchmDeut1977}, and `Proposed' to the data-aligned bins described in Section~\ref{ss:method2}. The number in brackets after a method name signifies the size of its buffer i.e.\ available working memory. }
  \label{f:ttp1_950}
\end{figure*}

Lastly, considering that this data set has stationary characteristics, we examined what we termed `time until accuracy'. Specifically, we define time until accuracy $t(\alpha)$ as the number of data points until the relative error of the estimate of an algorithm on a stream $s_1, s_2,\ldots,s_{n_s}$ permanently drops to at most $\alpha$:
\begin{align}
  t(\alpha) = \arg \max_{i=1,\ldots,n_s} \frac{|\hat{v}_q(i) - v_q(i)|}{v_q(i)} > \alpha,
\end{align}
where $v_q(i)$ is the true value of a quantile after the first $i$ data points have been seen, and $\hat{v}_q(i)$ an estimate of the quantile. The plots in Figure~\ref{f:ttp1_950} summarize the results obtained with different methods for accuracies of 0.01 (or 1\%), 0.05 (or 5\%), and 0.1 (or 10\%). The same trends observed thus far are apparent in these results as well.

\begin{figure*}[htb]
  \centering
  \subfloat[0.95 quantile]{\includegraphics[width=0.33\textwidth]{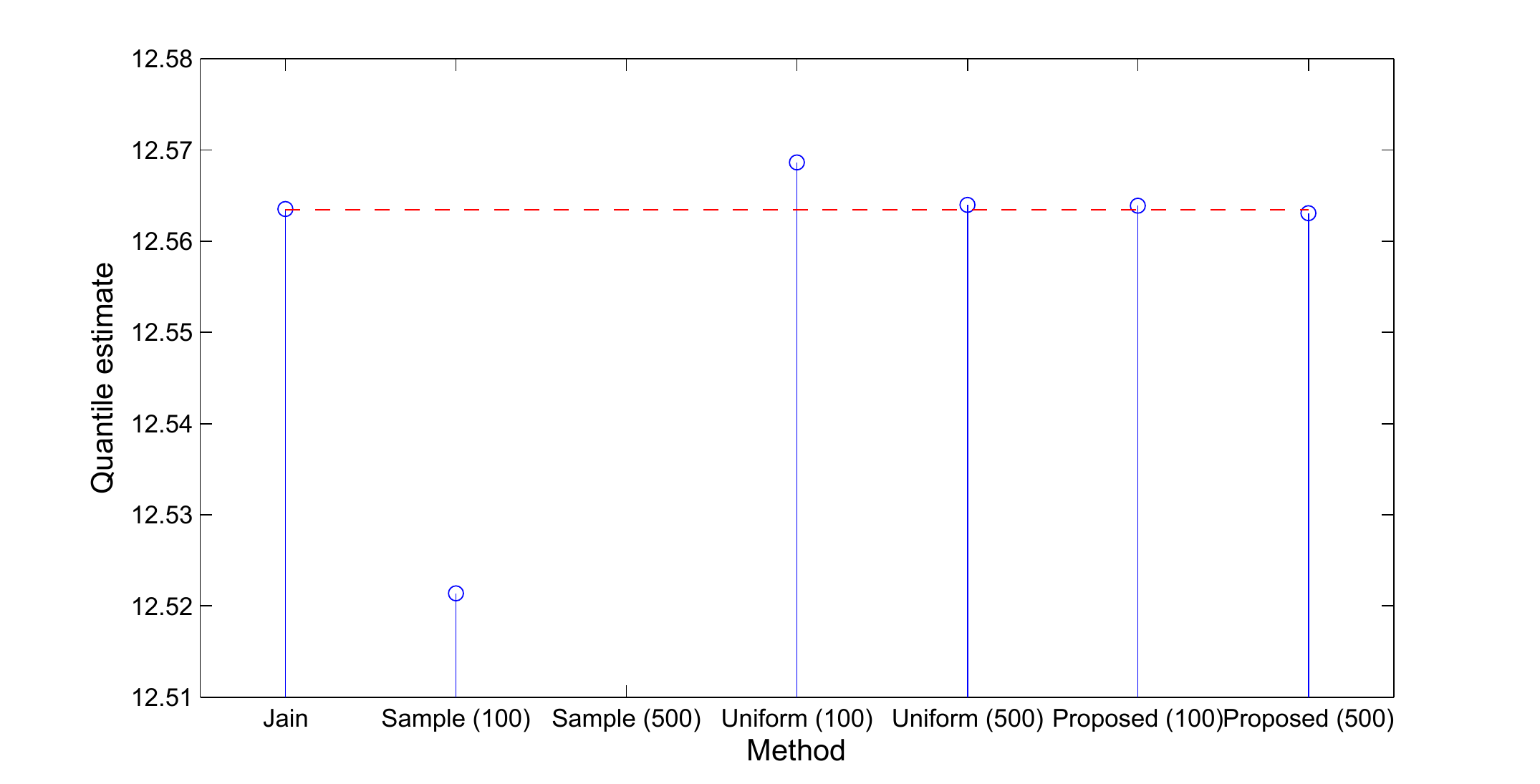}}
  \subfloat[0.99 quantile]{\includegraphics[width=0.33\textwidth]{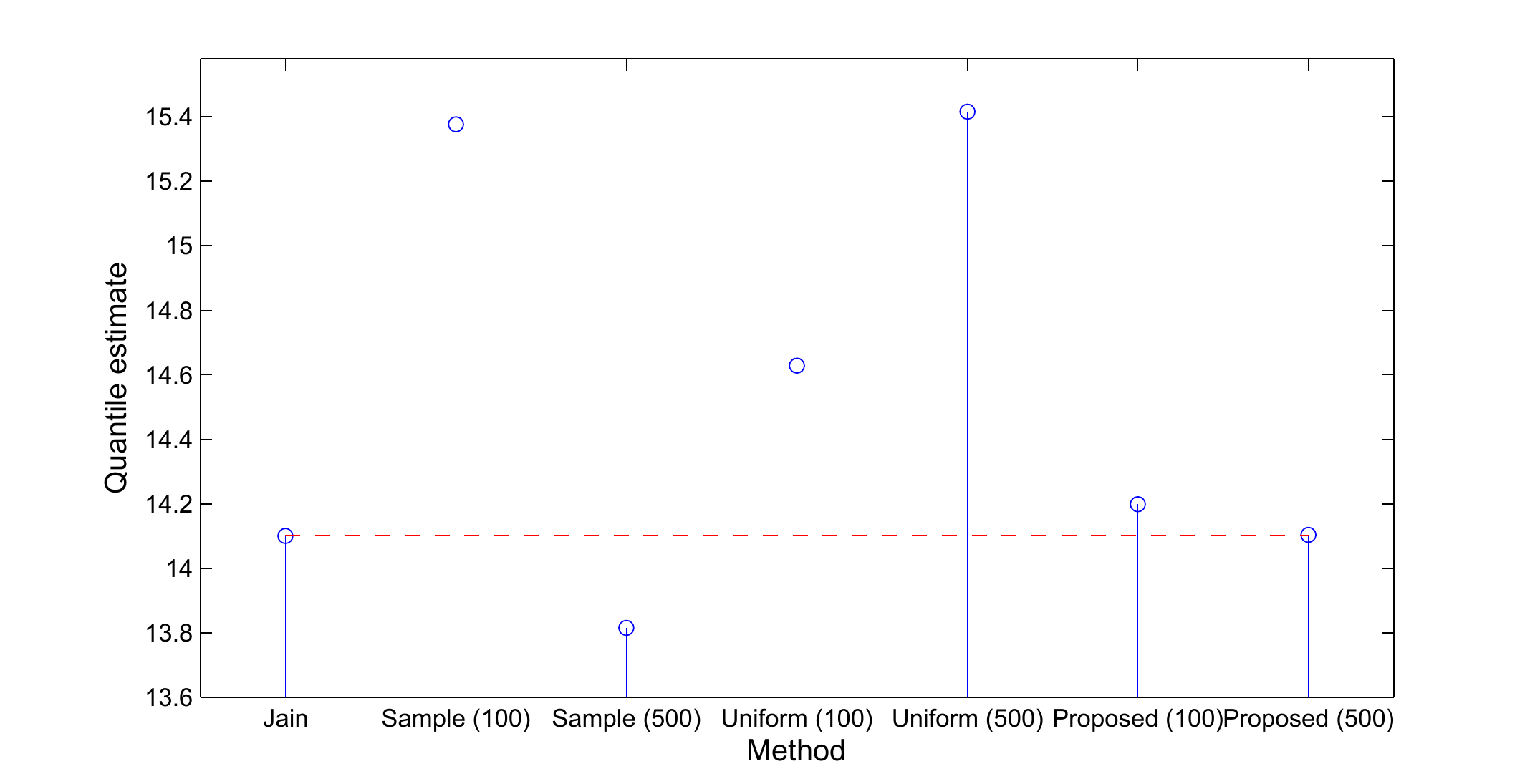}}
  \subfloat[0.995 quantile]{\includegraphics[width=0.33\textwidth]{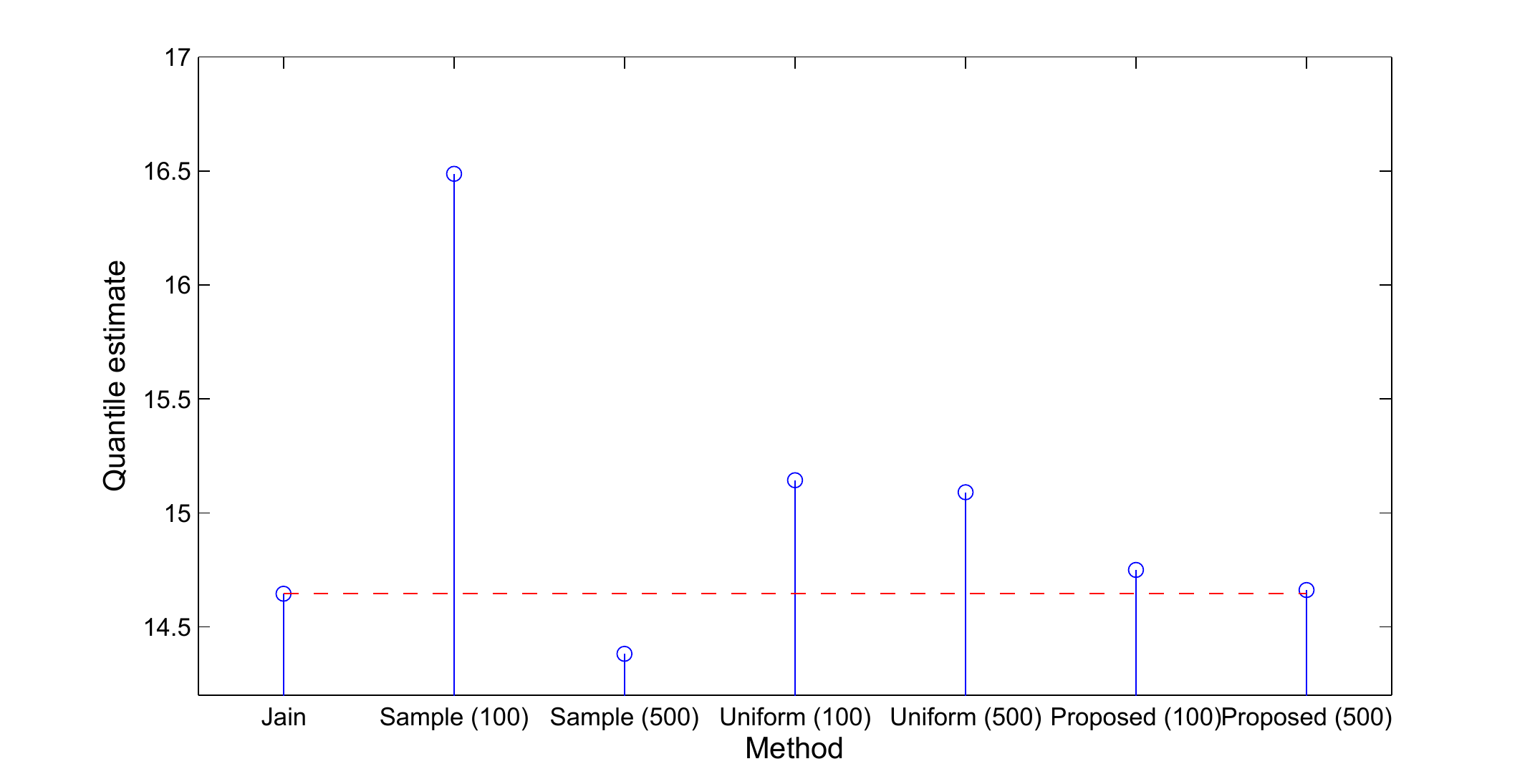}}
  \caption{ A comparison of different methods on the second synthetic data set used in this paper. Each datum of this stream (of 1,000,000 in total) was generated by flipping a fair coin and then depending on the outcome drawing the value either from the normal distribution $\mathcal{N}(5,1)$ or from $\mathcal{N}(10,4)$. Notice that data is not stationary in nature.  The label `Jain' refers to the $P^2$ algorithm of Jain and Chlamtac~\cite{JainChla1985}, `Sample' to the random sample based algorithm of Vitter~\cite{Vitt1985}, `Uniform' to the uniform adjustable histogram of Schmeiser and Deutsch~\cite{SchmDeut1977}, and `Proposed' to the data-aligned bins described in Section~\ref{ss:method2}. The number in brackets after a method name signifies the size of its buffer i.e.\ available working memory. The dotted red line shows the true quantile values. As in Figure~\ref{f:res_synth2} both of the proposed algorithms rapidly achieve high accuracy which is maintained throughout, resulting in running estimates indistinguishable by the naked eye at the scale of the plot. }
  \label{f:res_synth2}
\end{figure*}

\begin{figure}[htb]
  \centering
  \includegraphics[width=0.48\textwidth]{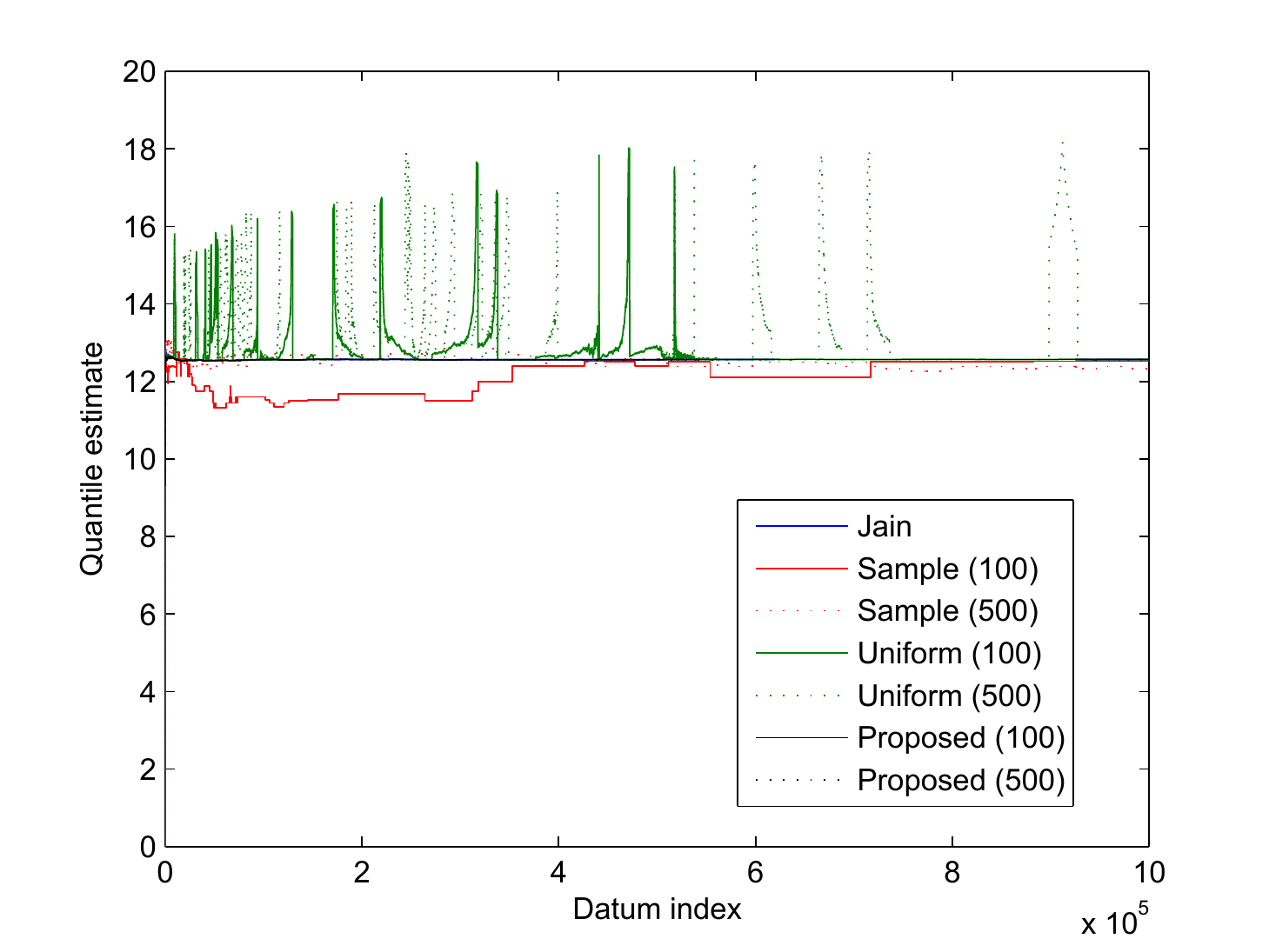}
  \caption{ Running estimate of the 0.95-quantile produced by different methods on our second (non-stationary) synthetic data set. The label `Jain' refers to the $P^2$ algorithm of Jain and Chlamtac~\cite{JainChla1985}, `Sample' to the random sample based algorithm of Vitter~\cite{Vitt1985}, `Uniform' to the uniform adjustable histogram of Schmeiser and Deutsch~\cite{SchmDeut1977}, and `Proposed' to the data-aligned bins described in Section~\ref{ss:method2}. The number in brackets after a method name signifies the size of its buffer i.e.\ available working memory.  }
  \label{f:res_synth2_plot}
\end{figure}

We now turn our attention to the second synthetic data set which, unlike the first one, does not exhibit stationary statistical properties. As before, we first summarized the estimates of three quantile values for different algorithms after all available data has been processed, as well as the running estimates. The results are summarized in respectively Figure~\ref{f:res_synth2} and Figure~\ref{f:res_synth2_plot}. Most of the conclusions which can be drawn from these mirror those already made on the first synthetic set. The proposed data-aligned bins algorithm consistently performed best and without any deterioration when the buffer size was reduced from 500 to 100. The uniform adjustable histogram of Schmeiser and Deutsch outperformed the sample-based algorithm of Vitter on average but again exhibited short-lived but large transient errors. The only major difference in comparison with the results obtained on the first data set is that in this case the simple method of Jain and Chlamtac performed extremely well (the corresponding running quantile estimate in Figure~\ref{f:res_synth2_plot} is indistinguishable from that of the proposed method on the scale shown).

\subsubsection{Real-world surveillance data}
\begin{figure*}[htb]
  \centering
  \subfloat[Proposed: interpolated bins]{\includegraphics[width=0.70\textwidth]{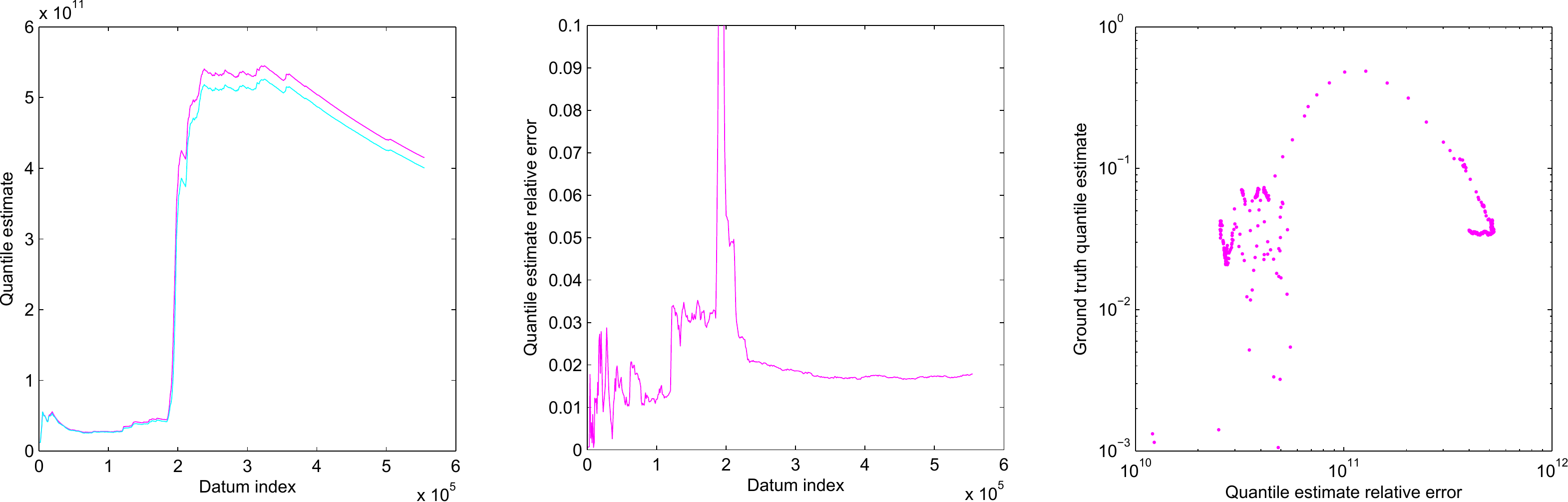}}\\
  \subfloat[Proposed: data-aligned bins]{\includegraphics[width=0.70\textwidth]{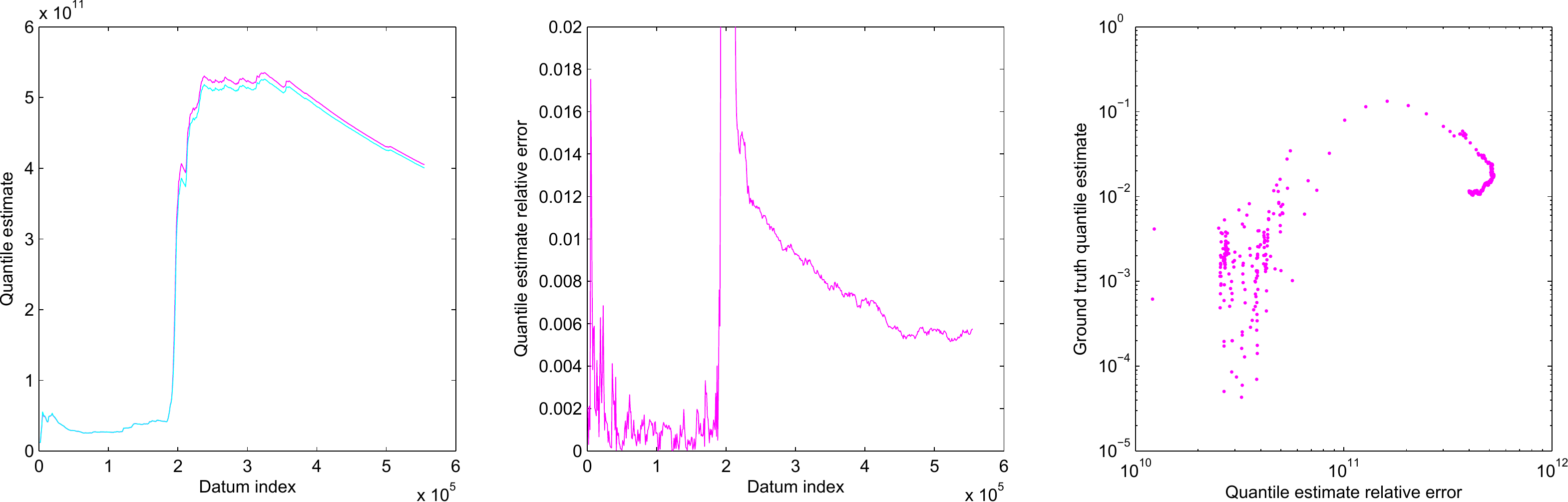}}\\
  \subfloat[$P^2$ algorithm \cite{JainChla1985}]{\includegraphics[width=0.70\textwidth]{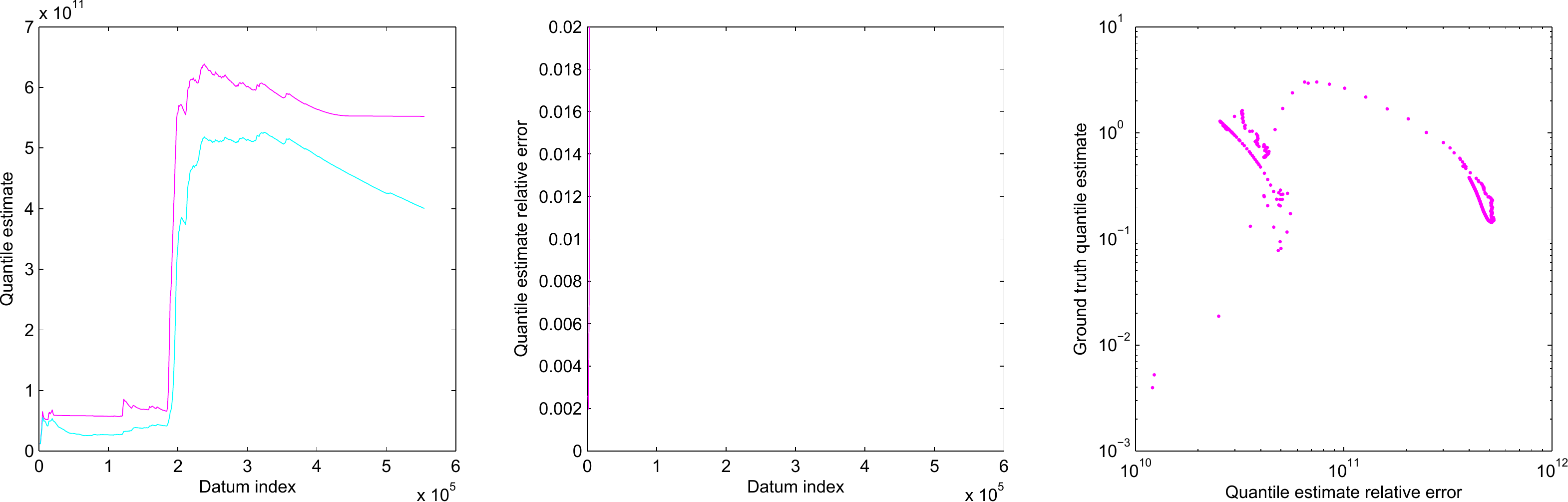}}\\
  \subfloat[Random sample \cite{Vitt1985}]{\includegraphics[width=0.70\textwidth]{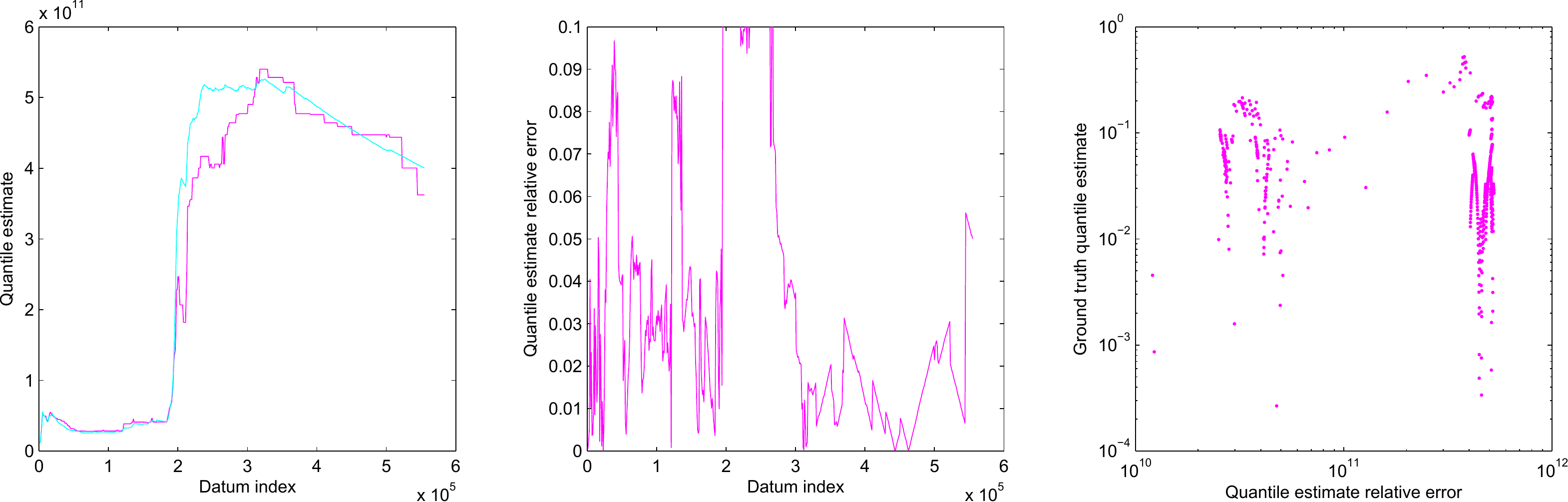}}\\
  \subfloat[Uniform histogram \cite{SchmDeut1977}]{\includegraphics[width=0.70\textwidth]{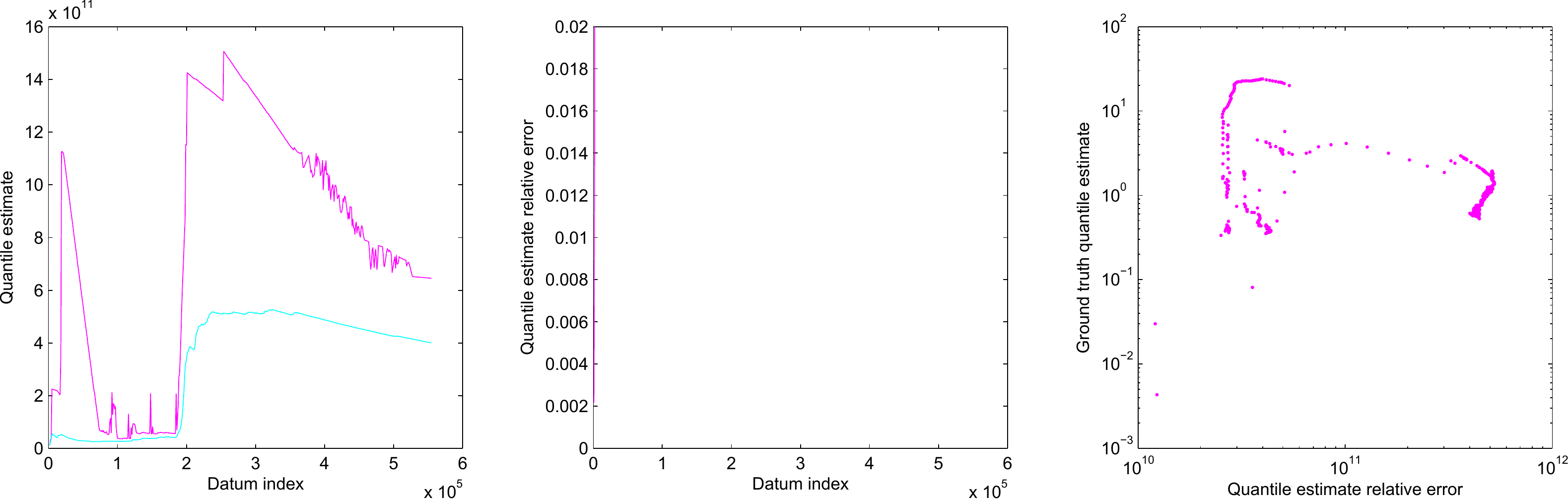}}
  \caption{ Running estimate of the 0.95-quantile on data stream~1.  A single row of plots corresponds to a particular method and shows, from left to right, the running quantile estimate of the algorithm (purple line) superimposed to the ground truth (cyan line), the error of the estimate relative to its true value, and the error of the estimate plotted as a function of the quantile value.  }
  \label{f:res095_x1}
\end{figure*}

\begin{figure*}[htb]
  \centering
  \subfloat[Proposed: interpolated bins]{\includegraphics[width=0.70\textwidth]{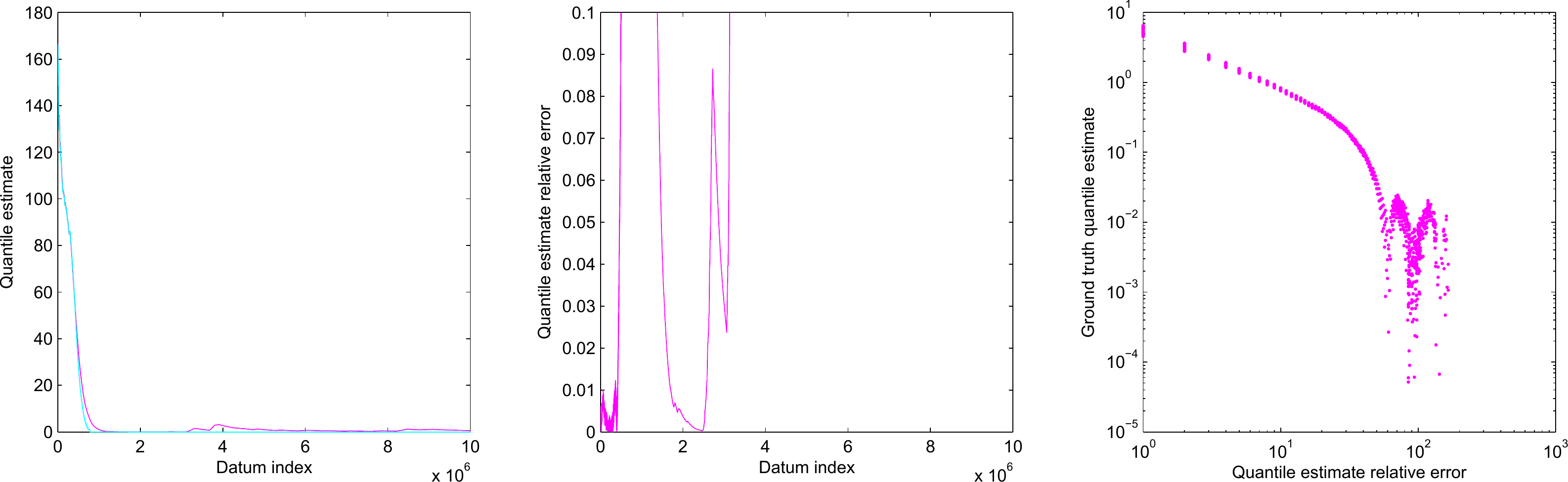}}\\
  \subfloat[Proposed: data-aligned bins]{\includegraphics[width=0.70\textwidth]{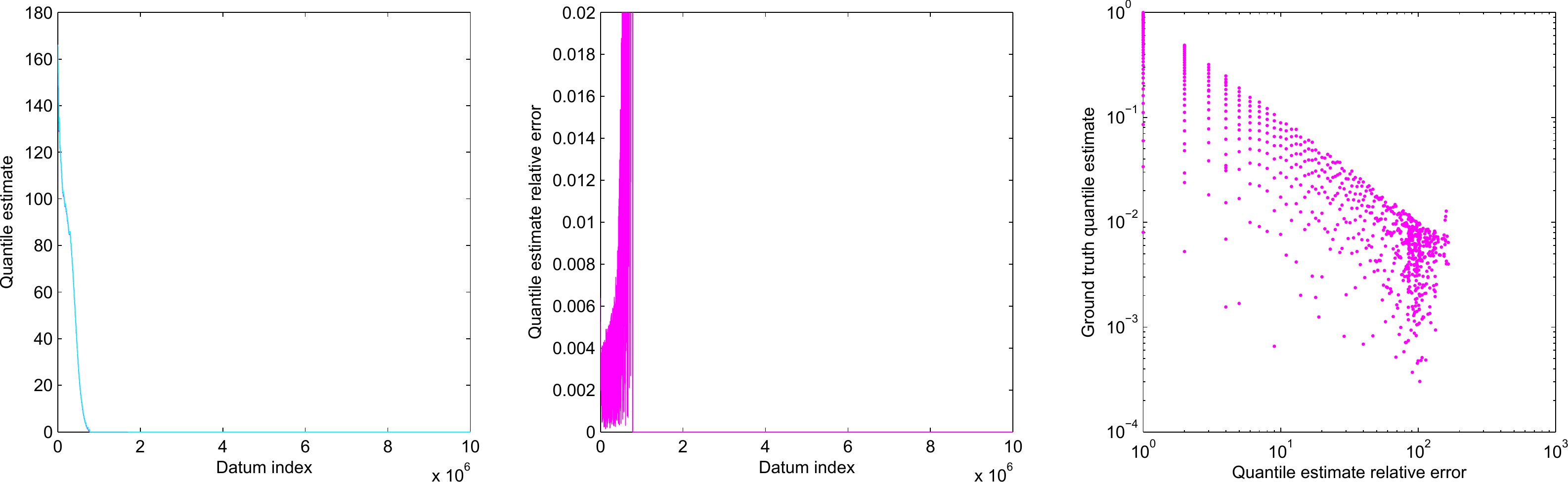}}\\
  \subfloat[$P^2$ algorithm \cite{JainChla1985}]{\includegraphics[width=0.70\textwidth]{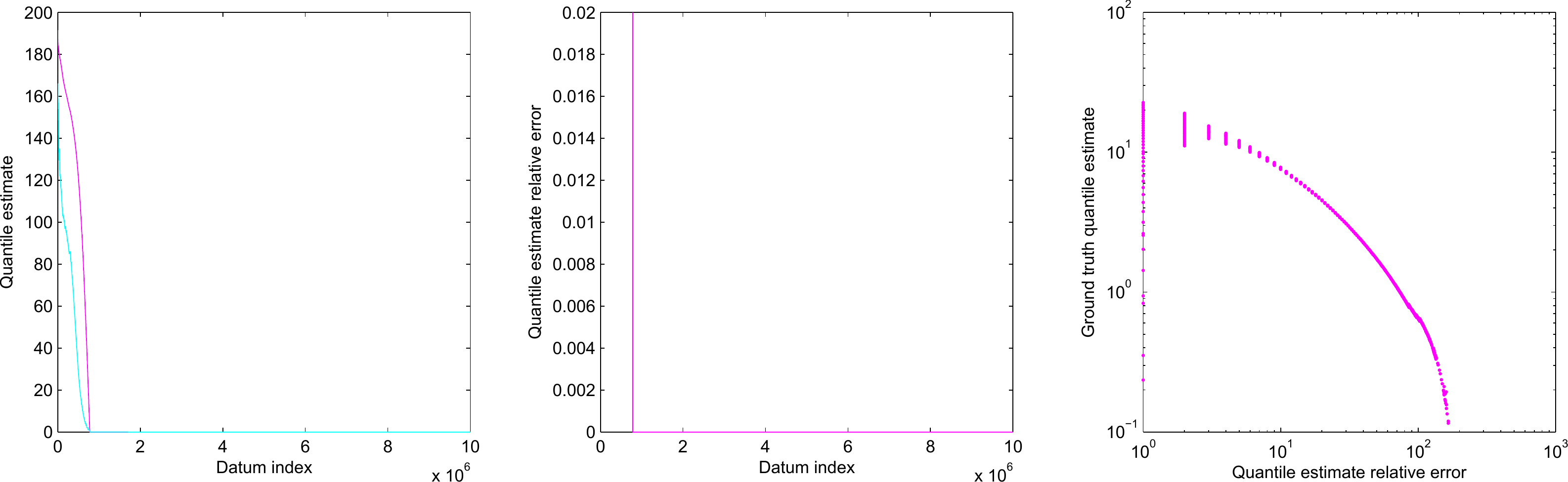}}\\
  \subfloat[Random sample \cite{Vitt1985}]{\includegraphics[width=0.70\textwidth]{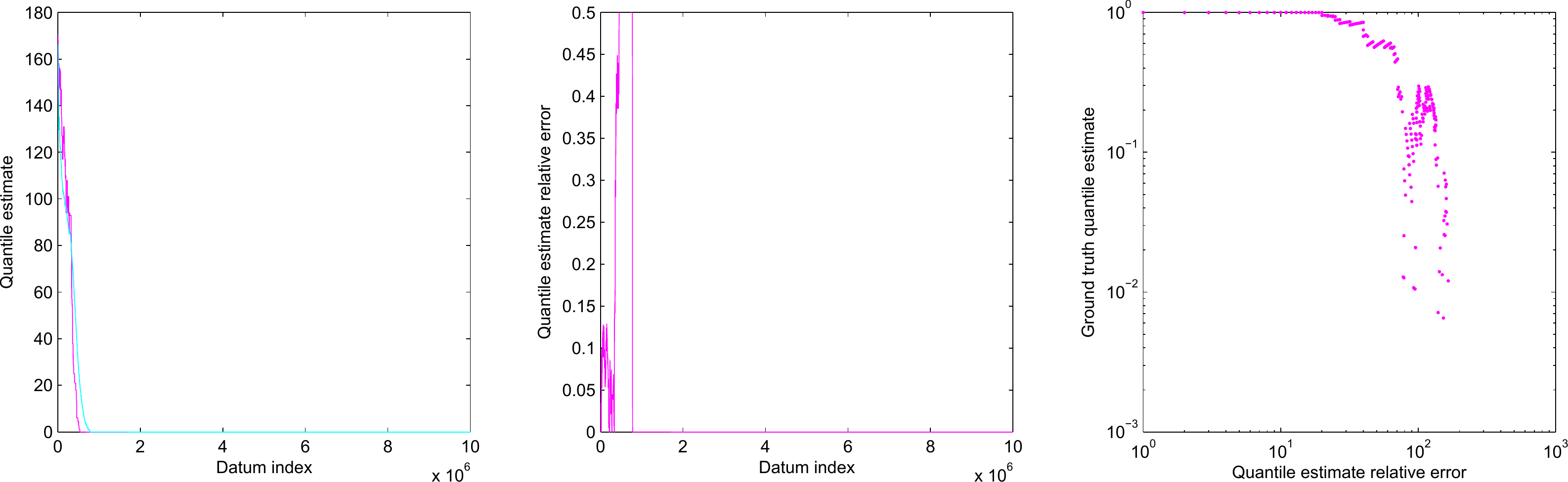}}\\
  \subfloat[Uniform histogram \cite{SchmDeut1977}]{\includegraphics[width=0.70\textwidth]{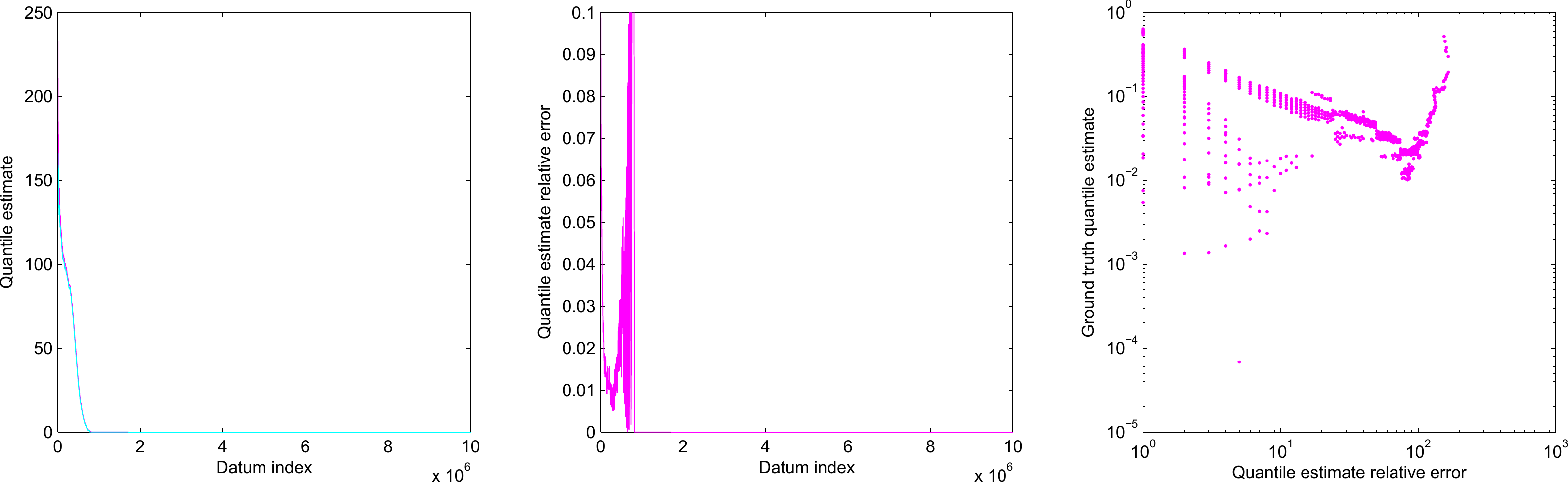}}
  \caption{ Running estimate of the 0.95-quantile on data stream~2. A single row of plots corresponds to a particular method and shows, from left to right, the running quantile estimate of the algorithm (purple line) superimposed to the ground truth (cyan line), the error of the estimate relative to its true value, and the error of the estimate plotted as a function of the quantile value. }
  \label{f:res095_x2}
\end{figure*}

\begin{figure*}[htb]
  \centering
  \subfloat[Proposed: interpolated bins]{\includegraphics[width=0.70\textwidth]{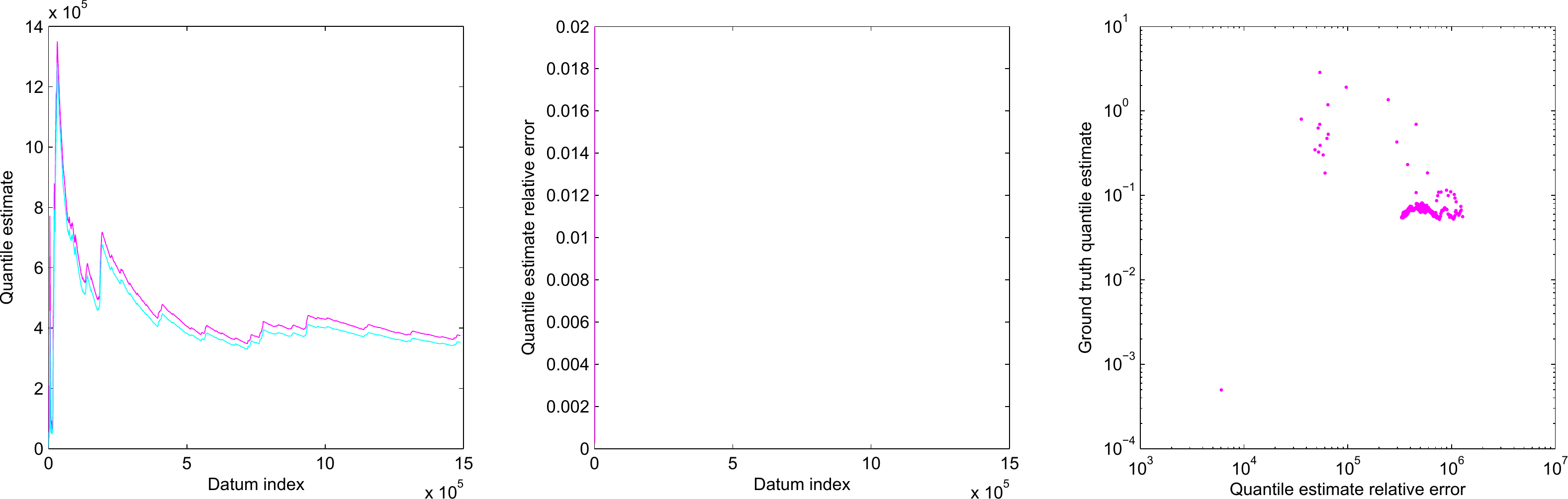}}\\
  \subfloat[Proposed: data-aligned bins]{\includegraphics[width=0.70\textwidth]{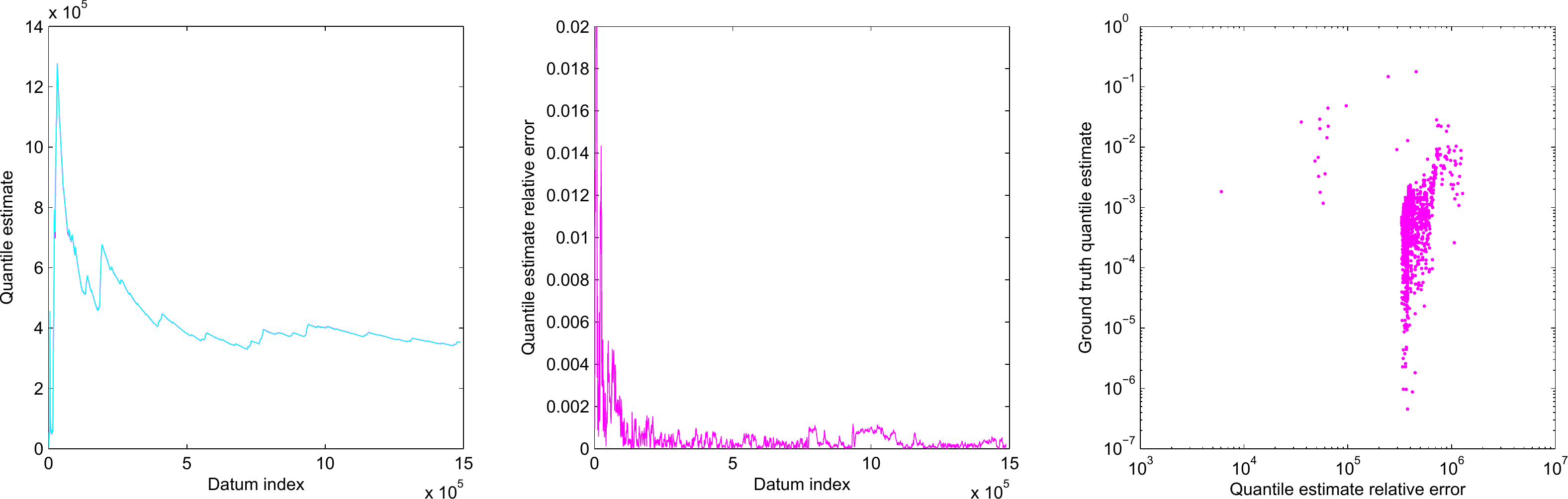}}\\
  \subfloat[$P^2$ algorithm \cite{JainChla1985}]{\includegraphics[width=0.70\textwidth]{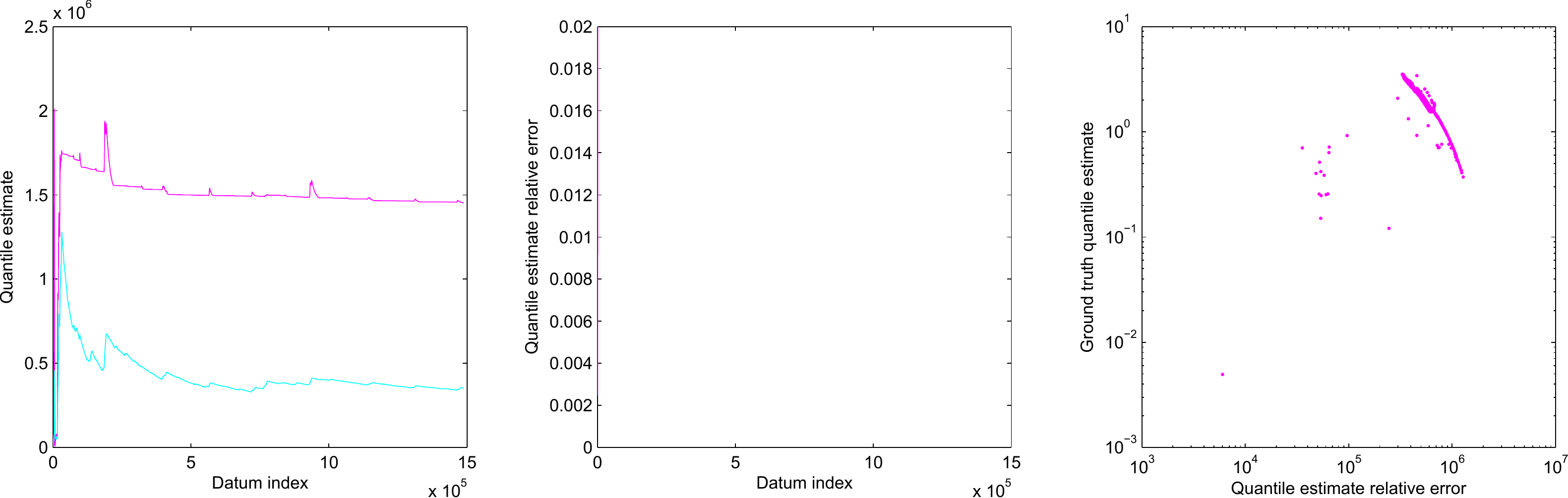}}\\
  \subfloat[Random sample \cite{Vitt1985}]{\includegraphics[width=0.70\textwidth]{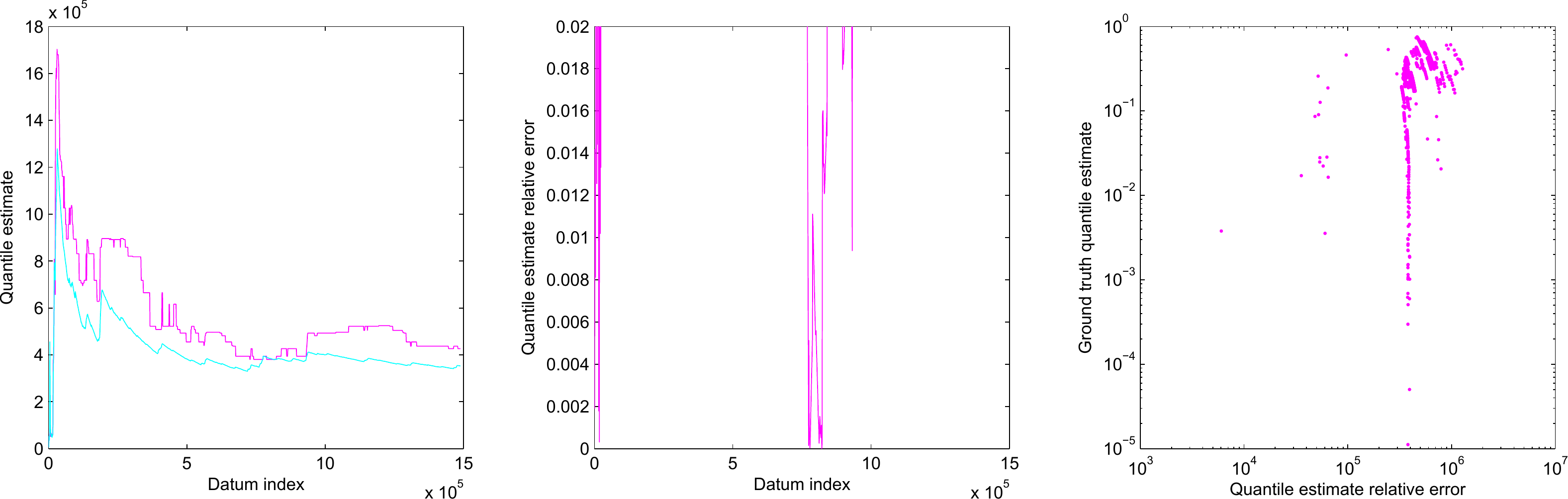}}\\
  \subfloat[Uniform histogram \cite{SchmDeut1977}]{\includegraphics[width=0.70\textwidth]{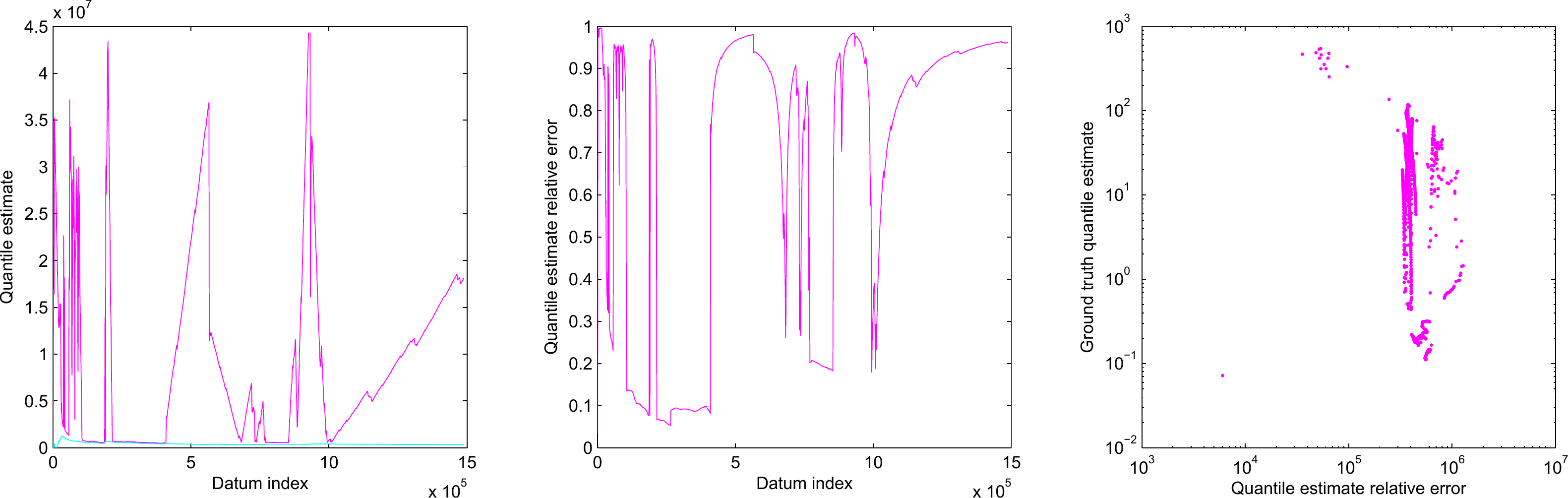}}
  \caption{ Running estimate of the 0.95-quantile on data stream~3. A single row of plots corresponds to a particular method and shows, from left to right, the running quantile estimate of the algorithm (purple line) superimposed to the ground truth (cyan line), the error of the estimate relative to its true value, and the error of the estimate plotted as a function of the quantile value. }
  \label{f:res095_x3}
\end{figure*}

Having gained some understanding of the behaviour of different algorithms in a setting in which input data was well understood and controlled, we applied them to data acquired by a real-world surveillance system. Representative results, obtained using the same number of bins $n=500$, for 0.95-quantile are shown in Figure~\ref{f:res095_x1}, Figure~\ref{f:res095_x2}, and Figure~\ref{f:res095_x3}, each corresponding to one of the three data streams. A single row of plots corresponds to a particular method and shows, from left to right, the running quantile estimate of the algorithm (purple line) superimposed to the ground truth (cyan line), the error of the estimate relative to its true value, and the error of the estimate plotted as a function of the quantile value. Firstly, compare the performances of the two proposed algorithms. In all cases and across time, the data-aligned bins algorithm produced a more reliable estimate. Thus, the argument put forward in Section~\ref{ss:method2} turned out to be correct -- despite the attempt of the interpolated bins algorithm to maintain exactly a maximum entropy approximation to the historical data distribution, the advantages of this approach are outweighed by the accumulation of errors caused by repeated interpolations. The data-aligned algorithm consistently exhibited outstanding performance on all three data sets, its estimate being virtually indistinguishable from the ground truth. This is witnessed and more easily appreciated by examining the plots showing its running relative error. In most cases the error was approximately 0.2\%; the only instances when the error would exceed this substantially are in the cases of very low quantile values (with even a tiny absolute error magnified in relative terms as in the case of stream~2), and transiently at times of sudden large change in the quantile value (as in the case of stream~1), quickly recovering thereafter.

\begin{figure}[htb]
  \centering
  \includegraphics[width=0.49\textwidth]{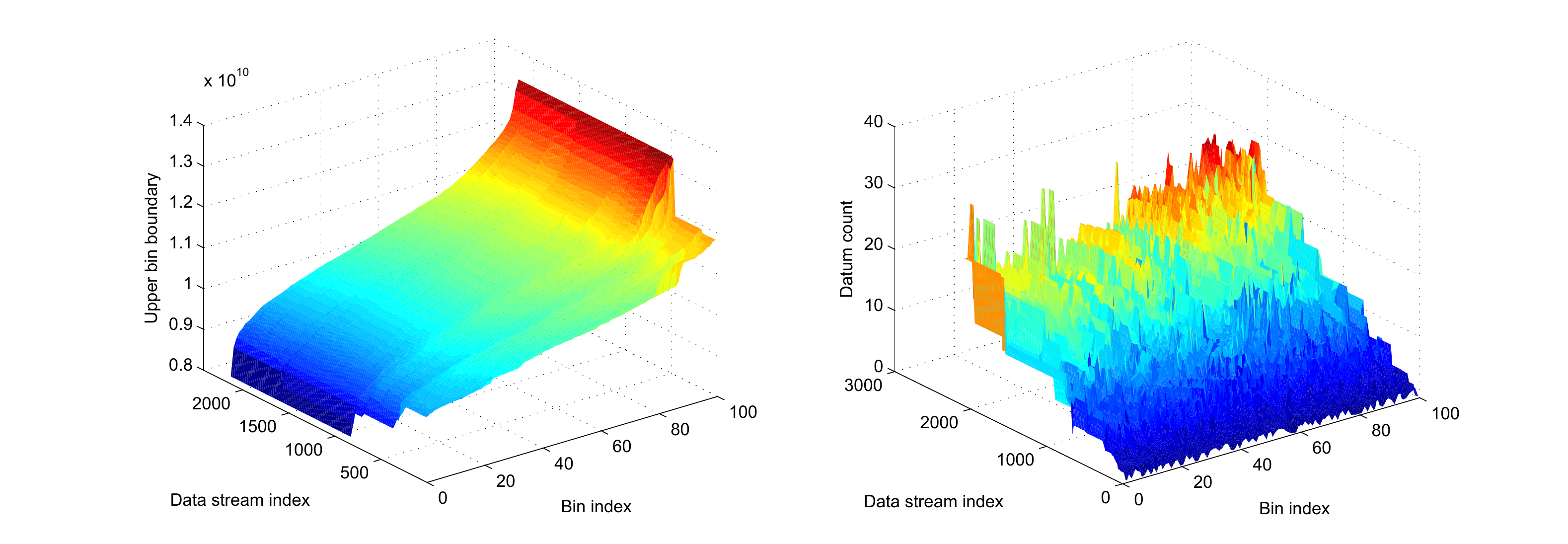}
  \caption{An illustration of a typical evolution of the histogram used in the proposed data-aligned bins algorithm. Shown on the left is the adaptation in the upper bin boundary values; on the right are the corresponding datum counts per bin. For this figure we used $n=100$ bins.}
  \label{f:binEvol}
%  \addtocounter{figure}{1}
\end{figure}

All of the algorithms from the literature performed significantly worse than both of the proposed methods. The assumption of stationary data statistics implicitly made in the $P^2$ algorithm and discussed in Section~\ref{ss:prev} is evident by its performance on stream~3. Following the initially good estimates when the true quantile value is relatively large, the algorithm is unable to adjust sufficiently to the changed data distribution and the decreasing quantile value. Across the three data sets, the random sample algorithm of Vitter~\cite{Vitt1985} overall performed best of the existing methods, never producing a grossly inaccurate estimate. Nonetheless its accuracy is far lower than that of the proposed algorithms, as easily seen by the naked eye and further witnessed by the corresponding plots of the relative error, with some tendency towards jittery and erratic behaviour. The adaptive histogram based algorithm of Schmeiser and Deutsch~\cite{SchmDeut1977} performed comparatively well on streams~2 and~3. On this account it may be surprising to observe its complete failure at producing a meaningful estimate in the case of stream~1. In fact the behaviour the algorithm exhibited on this data set is most useful in understanding the algorithm's failure modes. Notice at what points in time the estimate would shoot widely. If we take a look at Figure~\ref{f:stream1} it can be seen that in each case this behaviour coincides with the arrival of a datum which is much larger than any of the historical data (and thus the range of the histogram). What happens then is that in re-scaling the histogram by such a large factor, many of the existing bins get `squeezed' into only a single bin of the new histogram, resulting in a major loss of information. This is very much like what we observed previously on simple synthetic data in experiments in Section~\ref{sss:resSynth}. When this behaviour is contrasted with the performance of the algorithms we proposed in this paper, the importance of the maximum entropy principle as the foundational idea is easily appreciated; although our algorithms too readjust their bins upon the arrival of each new datum, the design of our histograms ensures that no major loss of information occurs regardless of the value of new data. Figure~\ref{f:binEvol} illustrates the evolution of bin boundaries and the corresponding bin counts in a run of our data-aligned histogram-based algorithm.

\begin{figure*}[htb]
  \centering
  \subfloat[Data stream~2, 0.99-quantile, 12 bins]{\includegraphics[width=0.70\textwidth]{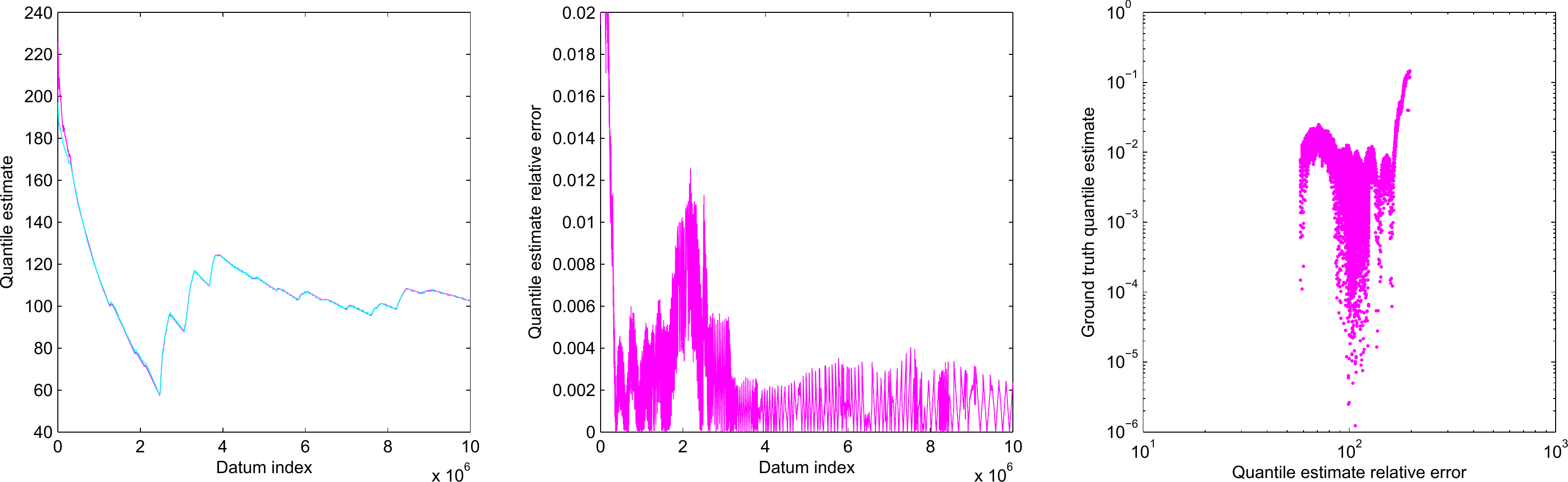}}\\
  \subfloat[Data stream~2, 0.99-quantile, 25 bins]{\includegraphics[width=0.70\textwidth]{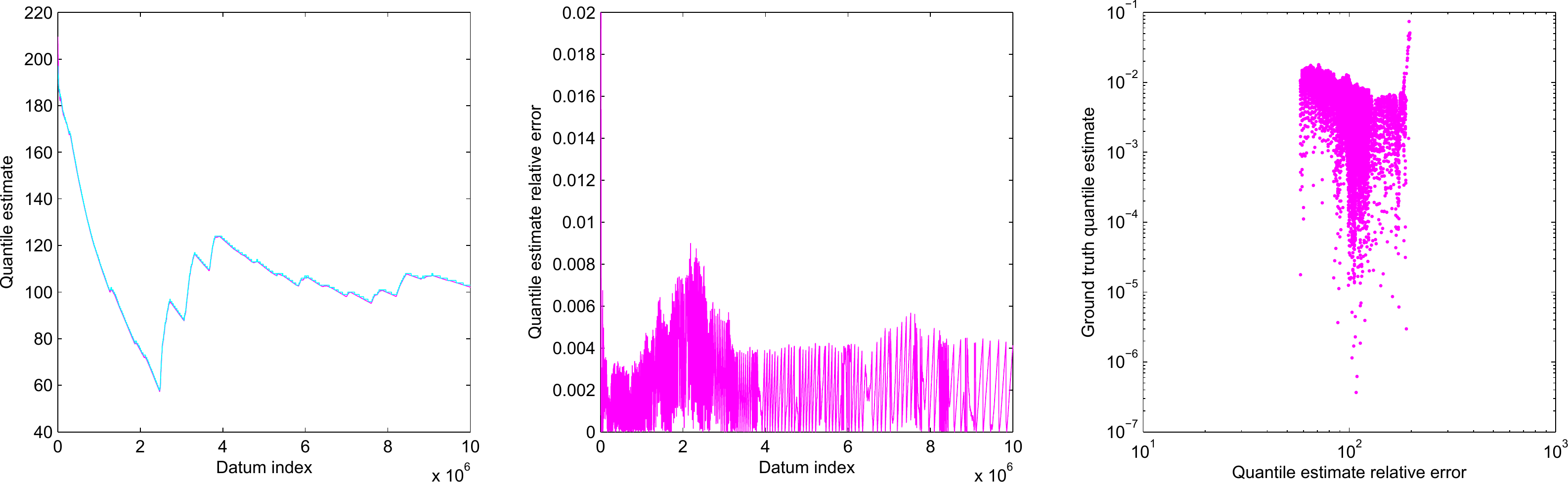}}\\
  \subfloat[Data stream~2, 0.99-quantile, 50 bins]{\includegraphics[width=0.70\textwidth]{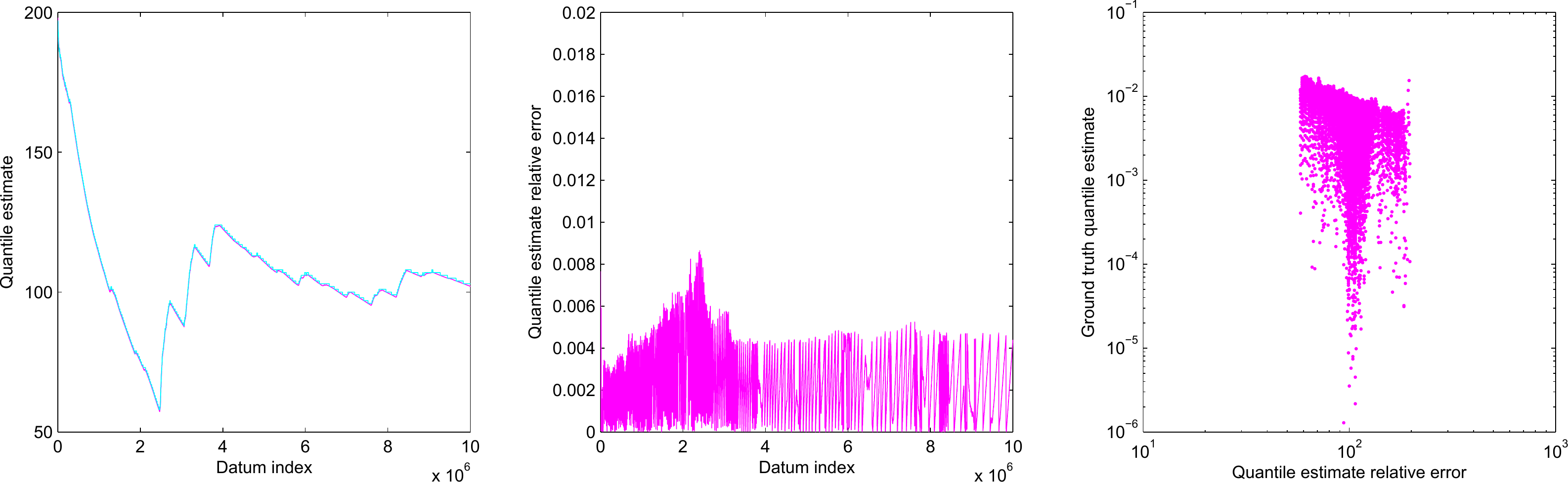}}\\
  \subfloat[Data stream~2, 0.99-quantile, 100 bins]{\includegraphics[width=0.70\textwidth]{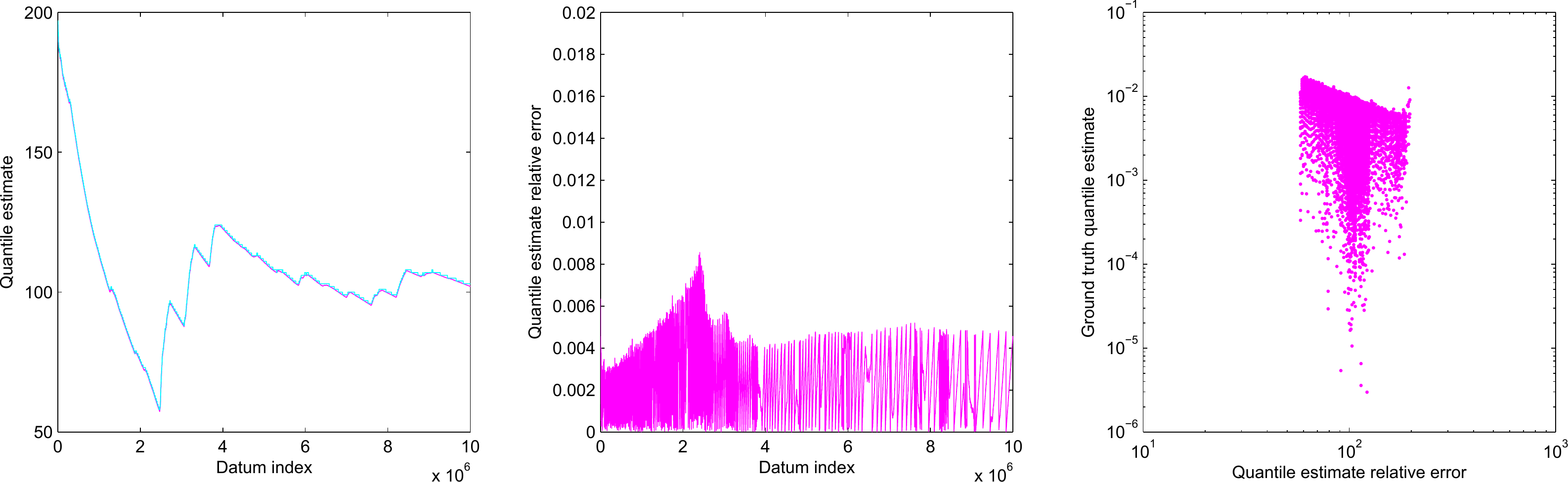}}\\
  \subfloat[Data stream~2, 0.99-quantile, 500 bins]{\includegraphics[width=0.70\textwidth]{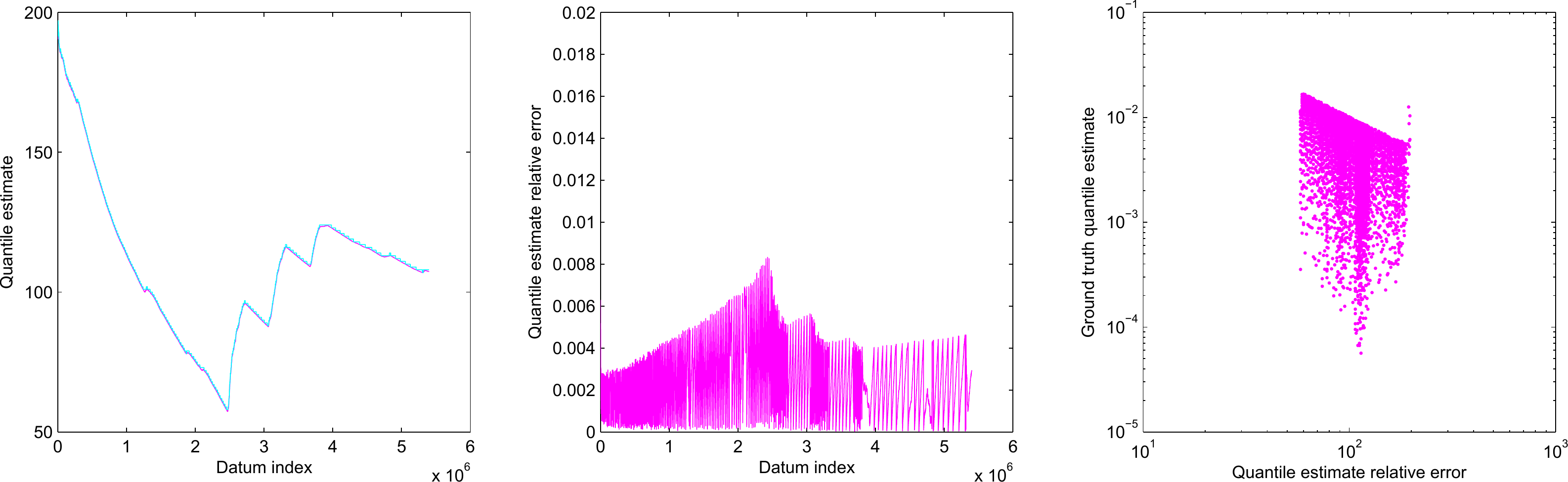}}
  \caption{ Running estimate of the 0.99-quantile on data stream~2 produced using our data-aligned adaptive histogram algorithm and different numbers of bins. A single row of plots corresponds to a particular method and shows, from left to right, the running quantile estimate of the algorithm (purple line) superimposed to the ground truth (cyan line), the error of the estimate relative to its true value, and the error of the estimate plotted as a function of the quantile value. Also see Table~\ref{t:res990}. }
  \label{f:reduction}
\end{figure*}

Considering the outstanding performance of our algorithms, and in particular the data-aligned histogram-based approach, we next sought to examine how this performance is affected by a gradual reduction of the working memory size. To make the task more challenging we sought to estimate the 0.99-quantile on the largest of our three data sets (stream~2). Our results are summarized in Table~\ref{t:res990}. This table shows the variation in the mean relative error as well as the largest absolute error of the quantile estimate for the proposed data-aligned histogram-based algorithm as the number of available bins is gradually decreased from 500 to 12. For all other methods, the reported result is for $n=500$ bins. It is remarkable to observe that the mean relative error of our algorithm does not decrease at all. The largest absolute error does increase, only a small amount as the number of bins is reduced from 500 to 50, and more substantially thereafter. This shows that our algorithm overall still produces excellent estimates with occasional and transient difficulties when there is a rapid change in the quantile value. Plots in Figure~\ref{f:reduction} corroborate this observation.

\begin{table}
  \small
  \centering
  \renewcommand{\arraystretch}{1.4}
  \caption{ A summary of the experimental results obtained on the real-world surveillance data set for the 0.99-quantile. Also see Figure~\ref{f:reduction}. }
  \vspace{0pt}
  \begin{tabular}{cc|cc}
  \Hline
  \multicolumn{2}{c}{Method} & Mean relative error & Absolute $L_\infty$ error\\
  \hline
   \multirow{5}{*}{\rotatebox{90}{Proposed}}
   \multirow{5}{*}{\rotatebox{90}{data-aligned}}
   \multirow{5}{*}{\rotatebox{90}{bins w/ bin no.}}
                                                             & 500 & 0.5\% & 2.43 \\
                                                             & 100 & 0.5\% & 2.45 \\
                                                             &  50 & 0.5\% & 3.01 \\
                                                             &  25 & 0.4\% & 14.48 \\
                                                             &  12 & 0.5\% & 28.83 \\
   \hline
    \multicolumn{2}{l|}{$P^2$ algorithm \cite{JainChla1985}} & 45.6\% & 112.61 \\
   \hline
    \multicolumn{2}{l|}{Random sample \cite{Vitt1985}}       & 17.5\% & 64.00 \\
    \hline
    \multicolumn{2}{l|}{Equispaced bins \cite{SchmDeut1977}} & 0.9\%  & 76.88 \\
  \Hline
  \end{tabular}
  \label{t:res990}
\end{table}

\section{Summary and conclusions}
In this paper we addressed the problem of estimating a desired quantile of a data set. Our goal was specifically to perform quantile estimation on a data stream when the available working memory is limited (constant), prohibiting the storage of all historical data. This problem is ubiquitous in computer vision and signal processing, and has been addressed by a number of researchers in the past. We show that a major shortcoming of the existing methods lies in their usually implicit assumption that the data is being generated by a stationary process. This assumption is invalidated in most practical applications, as we illustrate using real-world data extracted from surveillance videos.

Therefore we introduced two novel algorithms which deal with the described challenges effectively. Motivated by the observation that a consequence of non-stationarity is that the historical data distribution need not be representative of the future distribution and thus that a quantile value can change rapidly, we adopt a histogram-based representation to allow an adaptation to such unpredictable variability to take place. In contrast to the previous work which either distributes the bin boundaries equidistantly or uses \textit{ad hoc} adjustments, our idea was to maintain bins in a manner which maximizes the entropy of the corresponding estimate of the historical data distribution. The first method we described and which utilizes the stated principle readjusts by interpolation the locations of bin boundaries with every new incoming datum, attempting to maintain equiprobable bins. In contrast, our second algorithm constrains bin boundaries to the values of seen data. When a new datum arrives, a novel bin is created and the original number of bins restored by selecting the optimal (in the maximum entropy sense) merge of a pair of neighbouring bins.

The proposed algorithms were evaluated and compared against the existing alternatives described in the literature using three large data streams. This data was extracted from CCTV footage, not collected specifically for the purposes of this work, and represents specific motion characteristics over time which are used by semi-automatic surveillance analytics systems to alert to abnormalities in a scene. Our evaluation conclusively demonstrated a vastly superior performance of our algorithms, most notably the data-aligned bins algorithm. The highly non-stationary nature of our data was shown to cause major problems to the existing algorithms, often leading to grossly inaccurate quantile estimates; in contrast, our methods were virtually unaffected by it. What is more, our experiments demonstrate that the superior performance of our algorithms can be maintained effectively while drastically reducing the working memory size in comparison with the methods from the literature.

\balance

\clearpage

\bibliographystyle{ieeetran}
\bibliography{./my_bibliography}

\end{document}